\documentclass[journal]{IEEEtran}

\usepackage{subfigure}
\usepackage{enumerate}
\usepackage{multirow}
\usepackage{booktabs}
\usepackage{graphicx}
\usepackage{multirow}
\usepackage{amsmath}
\usepackage{balance}
\usepackage[table,xcdraw]{xcolor}
\usepackage{cite}

\ifCLASSINFOpdf
\else
\fi
\begin{document}

\title{Learning-Based Quality Assessment for Image Super-Resolution}


\author{Tiesong~Zhao,~\IEEEmembership{Senior~Member,~IEEE,}
	Yuting~Lin,
        Yiwen~Xu,~\IEEEmembership{Member,~IEEE,}
        Weiling~Chen,~\IEEEmembership{Member,~IEEE}
        and~Zhou~Wang,~\IEEEmembership{Fellow,~IEEE}
        % <-this % stops a space

%\thanks{}
\thanks{This work was supported by the National Natural Science Foundation of China (61901119). (\textit{Corresponding author: Weiling Chen.})}
\thanks{
T.  Zhao, Y. Lin, Y. Xu, W. Chen are with Fujian Key Lab for Intelligent Processing and Wireless Transmission of Media Information, Fuzhou University, Fuzhou 350116, China (E-mail: \{t.zhao, N181120063, xu\_yiwen, weiling.chen\}@fzu.edu.cn).}
\thanks{Z. Wang is with the Department of Electrical and Computer Engineering, University of Waterloo, Waterloo, ON, Canada (E-mail: Z.Wang@ece.uwaterloo.ca). }
}

\maketitle

\begin{abstract}
Image Super-Resolution (SR) techniques improve visual quality by enhancing the spatial resolution of images. Quality evaluation metrics play a critical role in comparing and optimizing SR algorithms, but current metrics achieve only limited success, largely due to the lack of large-scale quality databases, which are essential for learning accurate and robust SR quality metrics. In this work, we first build a large-scale SR image database using a novel semi-automatic labeling approach, which allows us to label a large number of images with manageable human workload. The resulting SR Image quality database with Semi-Automatic Ratings (SISAR), so far the largest of SR-IQA database, contains 8,400 images of 100 natural scenes. We train an end-to-end Deep Image SR Quality (DISQ) model by employing two-stream Deep Neural Networks (DNNs) for feature extraction, followed by a feature fusion network for quality prediction. Experimental results demonstrate that the proposed method outperforms state-of-the-art metrics and achieves promising generalization performance in cross-database tests. The SISAR database and DISQ model will be made publicly available to facilitate reproducible research.
%has arisen to enhance the existing low-resolution images and thus improve the visual quality at user-end. In this process, a quality evaluation metric can benefit the comparison and further optimizations of SR algorithms. However, the accurate SR evaluation is still a challenging task due to lack of large-scale database and deep models. To address this issue, this paper proposes a large-scale SR Image quality database with Semi-Automatic Ratings (SISAR) and a Deep Image SR Quality (DISQ) index. It is noted that the construction of large-scale database requires enormous image sources and incredible labelling workload. In this work, we observe a negatively exponential decay of visual quality when iteratively downsample and upscale a same image, which inspires us to establish a large-scale database with simplified labeling process.  The proposed SISAR database consists of over 8,400 scored images in 100 different natural scenes, with a promisingly high correlation to one-by-one labelled scores. Based on this database, we propose an end-to-end deep network for SR image quality assessment with reduced-reference. The proposed DISQ model employs a two-stream deep neural networks to fully exploit the features of images before and after SR. Finally, the two networks are connected by an optimized feature fusion step to regress image qualities effectively. Experimental results demonstrate that the proposed method outperforms other related metrics and achieves sufficient generalization performance.
\end{abstract}


% \fi
\IEEEpeerreviewmaketitle

\section{Introduction}
\begin{figure}[t]
  \centering
%  \subfigure[LR]{
 % \includegraphics[width=1.8cm]{fig/animal9/LR.jpg}
  %}
  
    \subfigure[BICUBIC\_2]{
  \includegraphics[width=1.8cm]{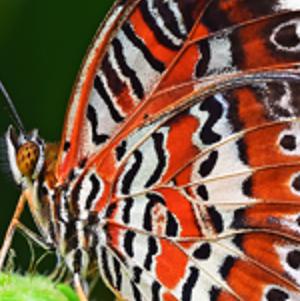}
  }
      \subfigure[RLLR\_2]{
  \includegraphics[width=1.8cm]{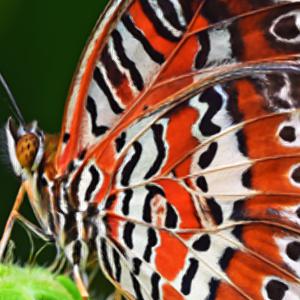}
  }
    \subfigure[SRCNN\_2]{
  \includegraphics[width=1.8cm]{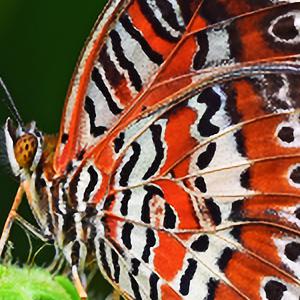}
  }
    \subfigure[VDSR\_2]{
  \includegraphics[width=1.8cm]{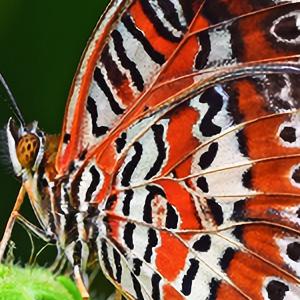}
  }
 
     \subfigure[BICUBIC\_3]{
  \includegraphics[width=1.8cm]{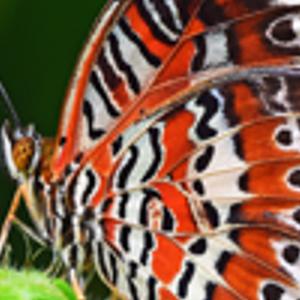}
  }
      \subfigure[RLLR\_3]{
  \includegraphics[width=1.8cm]{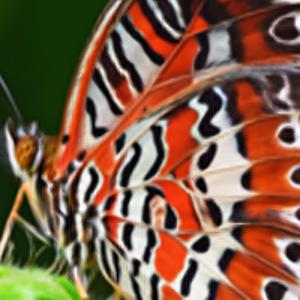}
  }
    \subfigure[SRCNN\_3]{
  \includegraphics[width=1.8cm]{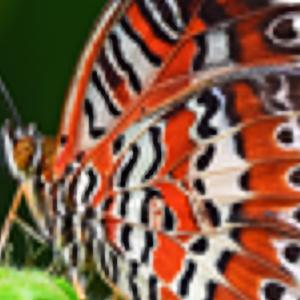}
  }
    \subfigure[VDSR\_3]{
  \includegraphics[width=1.8cm]{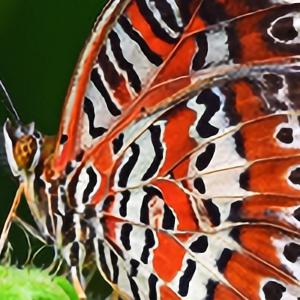}
  }
  
      \subfigure[BICUBIC\_4]{
  \includegraphics[width=1.8cm]{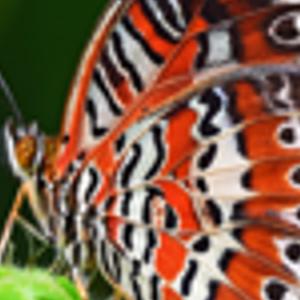}
  }
        \subfigure[RLLR\_4]{
  \includegraphics[width=1.8cm]{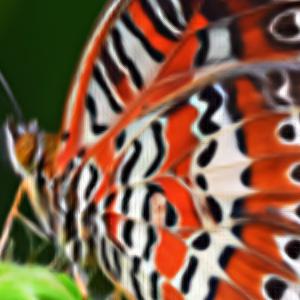}
  }
    \subfigure[SRCNN\_4]{
  \includegraphics[width=1.8cm]{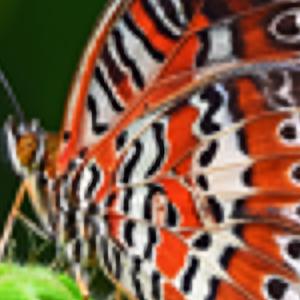}
  }
    \subfigure[VDSR\_4]{
  \includegraphics[width=1.8cm]{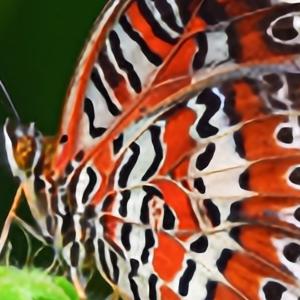}
  }
  \caption{Examples of HR images generated by SR. (a-d), (e-h) and (i-l) are reconstructed HR images via BICUBIC \cite{Hou1979}, RLLR \cite{Liu2012}, SRCNN \cite{Dong2016} and VDSR \cite{Kim2016} algorithms with scaling factors of 2, 3 and 4 respectively. Quality variations are observed across both scaling factor and SR methods.}
  \label{fig:SR}
\end{figure}

\IEEEPARstart{W}{ith} the rapid development of high-definition displays, the demand for high resolution image/video content has been increasing rapidly. To improve the user-end visual experience, image Super-Resolution (SR) technique is developed to interpolate High-Resolution (HR) images from their Low-Resolution (LR) references. Examples of these interpolated HR images can be seen in Fig. \ref{fig:SR}. The past two decades have witnessed a booming of image SR algorithms with widespread applications including medical image processing, video surveillance, remote sensing, and face recognition, among many others. In the SR process, an Image Quality Assessment (IQA) metric is critical as both a performance indicator and a guidance for further improvement. However, the mainstream IQA methods, such as Peak Signal-to-Noise Ratio (PSNR) and Structural SIMilarity (SSIM) index \cite{Wang2004}, do not have high correlation with subjective opinions in image SR \cite{Wang2020}.

Existing IQA methods can be divided into two categories: subjective tests by human observers and objective models by automatic algorithms. Between them, the more reliable way is the subjective test, since human users are the ultimate viewers of images. Several SR methods assess their performances using small-scale subjective tests \cite{Mehdi2017, WangRecovering2018}. However subjective test is time-consuming and hard to be embedded into practical applications. Instead, it is often utilized to construct databases, which serve as standard testing sets to compare objective models. 

Numerous objective IQA models have also been proposed. According to the availability of reference, objective models can be classified into full-reference, reduced-reference and no-reference IQA models. In a typical image SR problem, the perfect quality HR image is unavailable as a reference, therefore, full-reference models are generally inapplicable. In addition,  SR algorithms often introduce mixed impairments, including blurring, ringing and aliasing artifacts, which are not well measured by the existing methods \cite{Ma2016,Tang2019}. Therefore, conventional general-purpose reduced-reference and no-reference IQA models should be redesigned specifically for image SR problem. 

The key challenge in SR-IQA is how to effectively learn distinguishing feature representations of various LR and HR images and then map the features to image quality prediction. Convolutional Neural Network (CNN) has shown its advantage in IQA with promising successes in recent years \cite{Kang2014, Yang2019,Yan2019}. Compared with other data-driven models, CNN is capable of learning features and making quality predictors jointly in an end-to-end manner. Despite its superiority, CNN has not yet been well exploited in SR-IQA, for which most methods are not based on deep learning \cite{Chen2018, Yeganeh2015, Fang2016, Tang2019, Ma2016}.  In this work, we propose a CNN-based SR-IQA algorithm. Considering the importance of reference information \cite{Bosse2018}, we employ CNNs to extract the features from both images before and after the SR process. A two-stream CNN architecture is thereby designed to simultaneously take the test HR and the corresponding LR reference images, followed by feature fusion and quality prediction of the HR image.

The performance of deep learning based models relies heavily on the quality and quantity of training images. Although several benchmark databases of image SR qualities have been constructed \cite{Yang2014, Yeganeh2015, Ma2016, Wang2017, Zhou2019}, they are limited in their sample sizes. The largest database contains only 1,620 HR images. A larger image SR database is thus imperative, for which the biggest challenge is in the enormous workload of human labeling. In this work, we observe a negatively exponential decay behavior of subjective scores after iterative downsample-and-SR processing of natural images. This helps us develop a large-scale database with reduced human labeling workload. Experiments on randomly selected samples demonstrate the high accuracy of this database. Based on this database, we are able to develop and train the two-stream deep network that predicts the quality of SR images with a high correlation to human scores.

Our major contributions are as follows:
\begin{enumerate}[1)]
%\vspace{3mm}
\item Developed so far the largest IQA database for image SR with a novel semi-automatic labeling method, which greatly reduced the workload of human labeling.
\vspace{0.5mm}
\item Proposed a two-stream deep network that incorporates the available LR image into the quality evaluation of HR image. It results in an end-to-end model that jointly learns perceptually consistent features from the two images and a quality predictor.  
\vspace{0.5mm}
\item Designed a feature fusion method that combines the two-stream deep CNN, leading to superior performance against state-of-the-art quality prediction.
%\vspace{3mm}
\end{enumerate}

The remaining of this paper is organized as follows. Section \uppercase\expandafter{\romannumeral2} introduces the related work. Section \uppercase\expandafter{\romannumeral3} explains the methodology and process to construct the proposed large-scale database. Section \uppercase\expandafter{\romannumeral4} elaborates the proposed deep network, including network architecture, feature fusion method, and model training. Section \uppercase\expandafter{\romannumeral5} provides the experimental results. Finally, the paper is concluded in Section \uppercase\expandafter{\romannumeral6}.

\begin{figure*}[t]
	\centering
	\subfigure[ANIMALS: 0.9882]{
		\begin{minipage}{0.19\linewidth}
			\includegraphics[width=2.8cm]{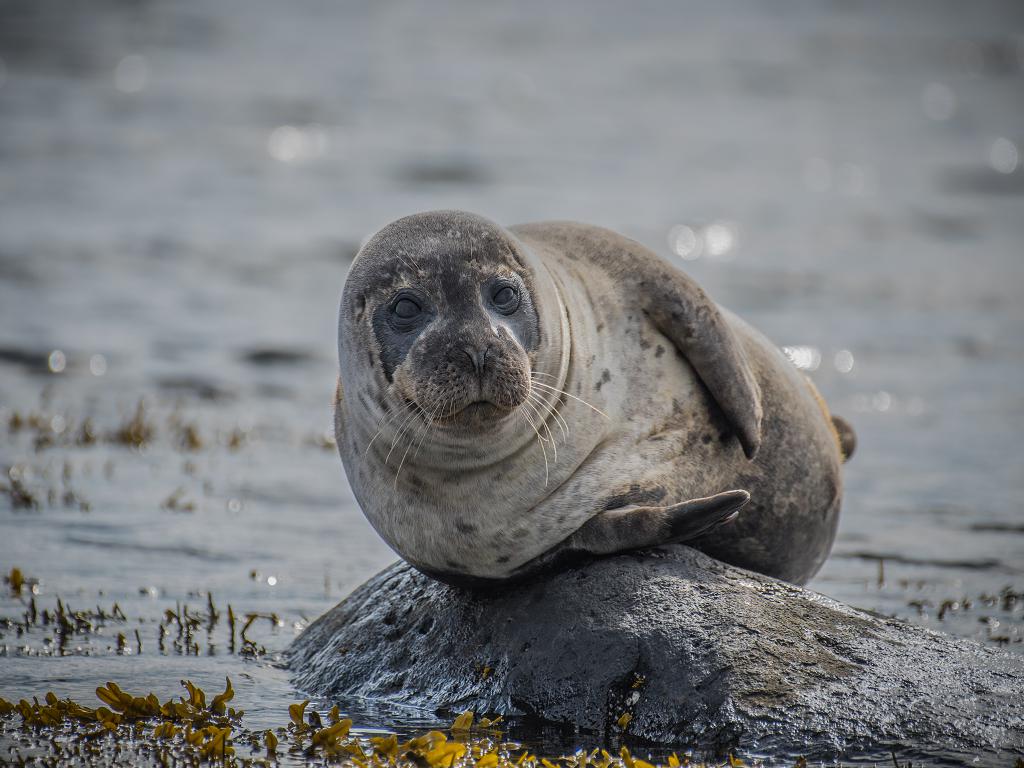}\vspace{4pt}
			\includegraphics[width=2.85cm]{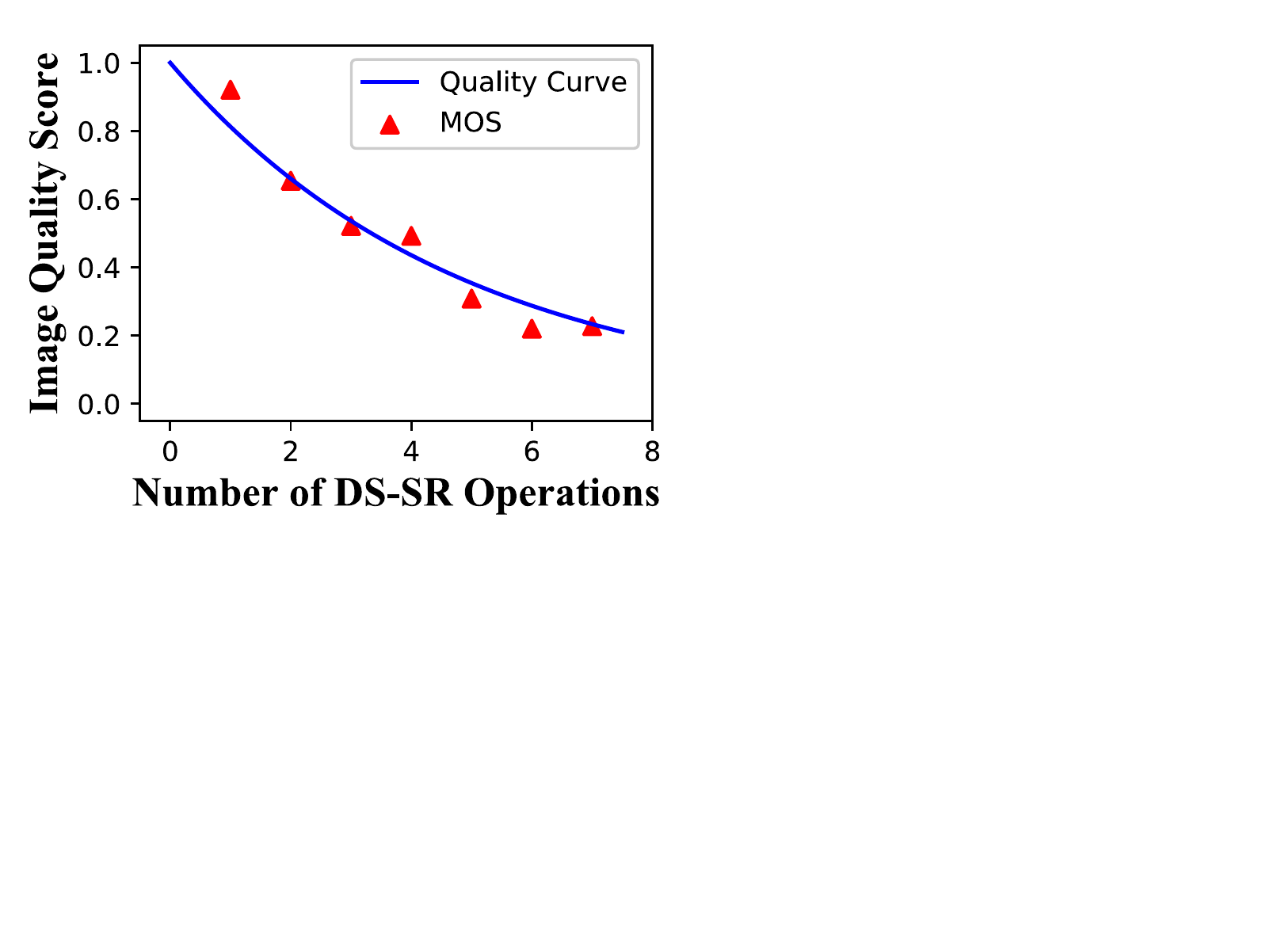}
	\end{minipage}}\hspace{-7.5mm}
	\subfigure[BUILDINGS: 0.9875]{
		\begin{minipage}{0.19\linewidth}
			\includegraphics[width=2.8cm]{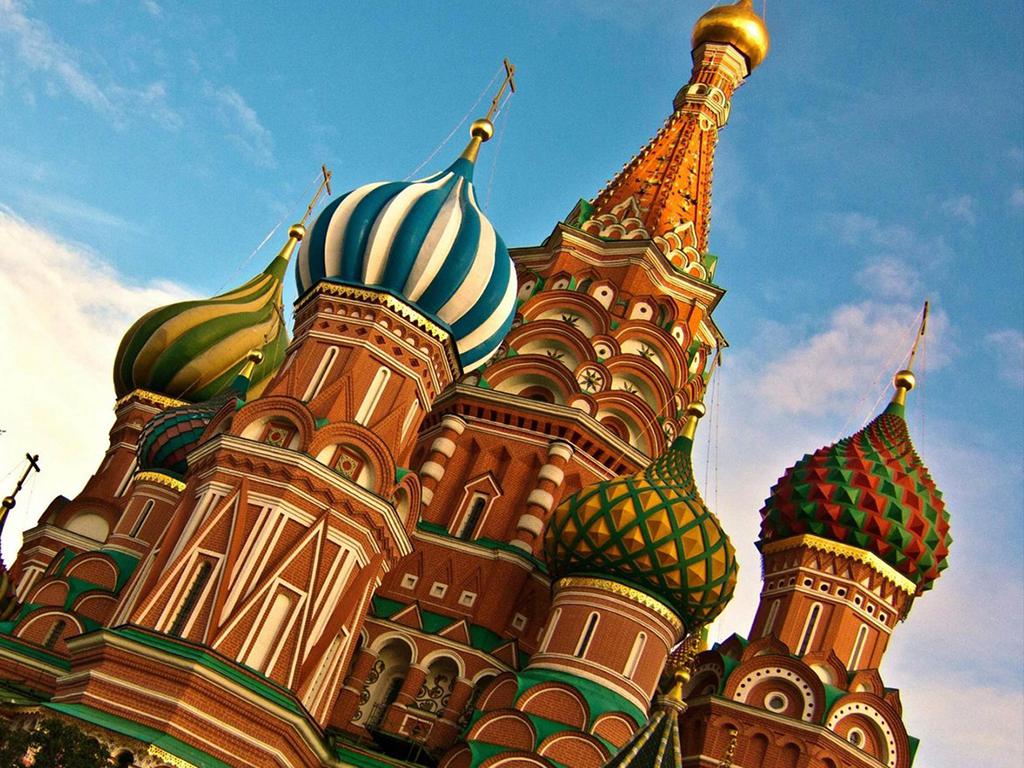}\vspace{4pt}
			\includegraphics[width=2.85cm]{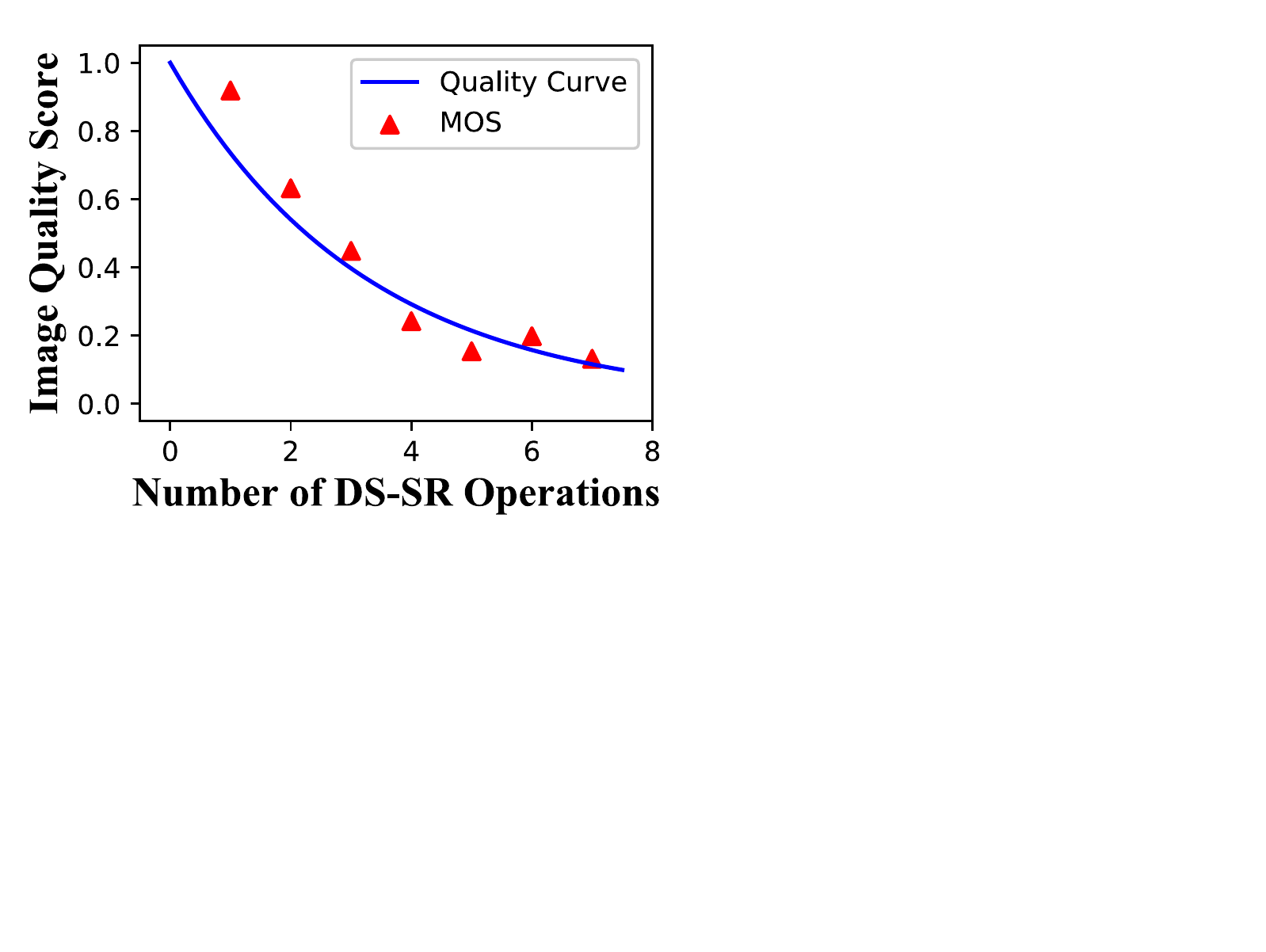}
	\end{minipage}}\hspace{-7.5mm}
	\subfigure[HUMANS: 0.9923]{
		\begin{minipage}{0.19\linewidth}
			\includegraphics[width=2.8cm]{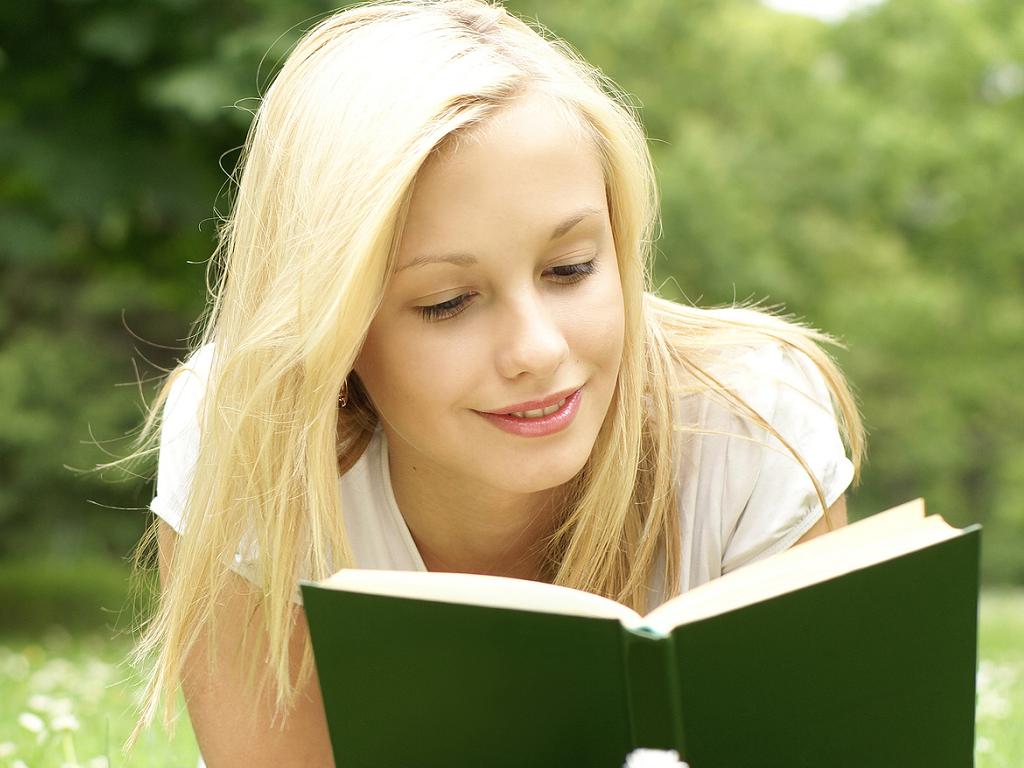}\vspace{4pt}
			\includegraphics[width=2.85cm]{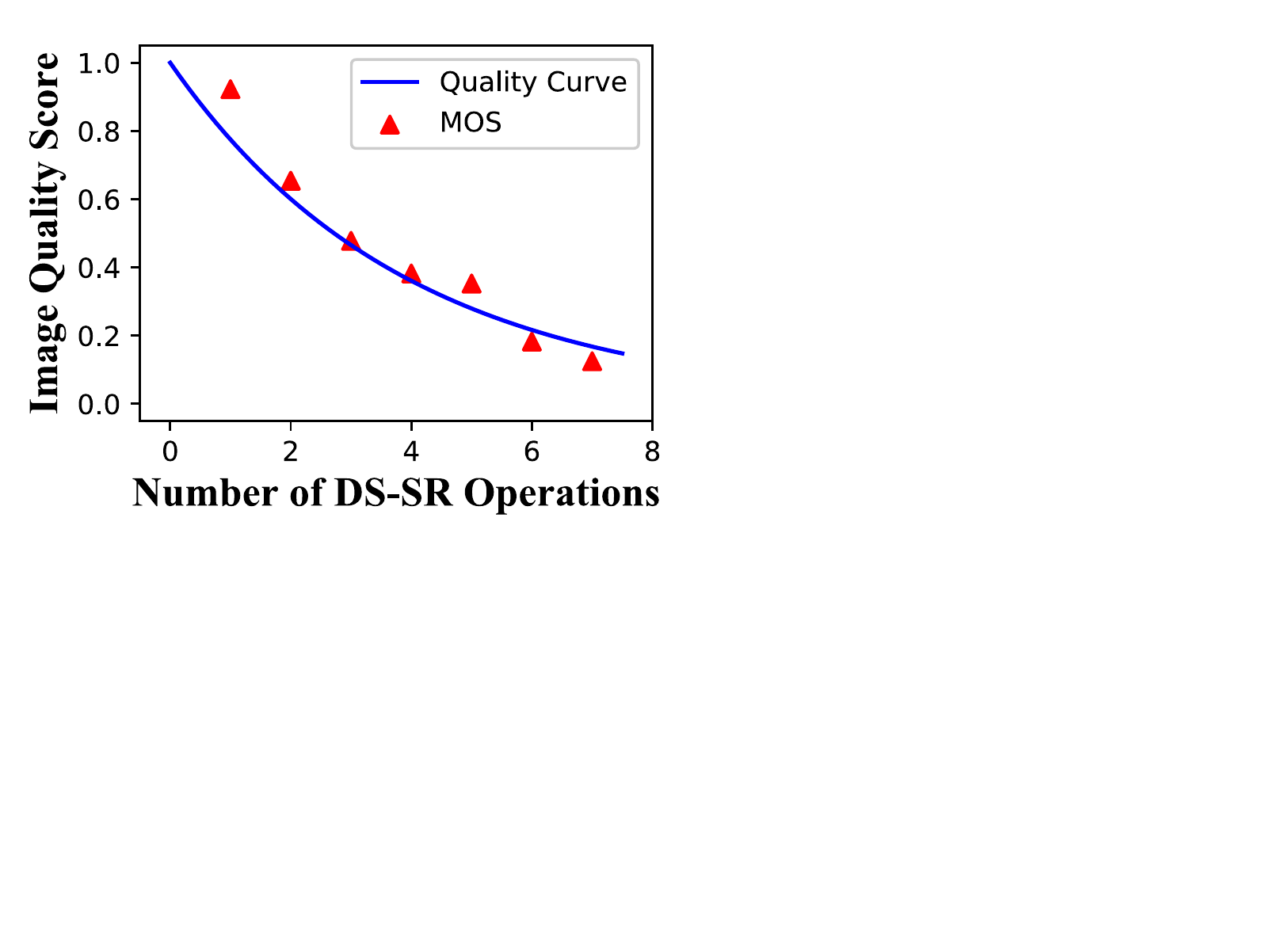}
	\end{minipage}}\hspace{-7.5mm}
	\subfigure[SPORTS: 0.9932]{
		\begin{minipage}{0.19\linewidth}
			\includegraphics[width=2.8cm]{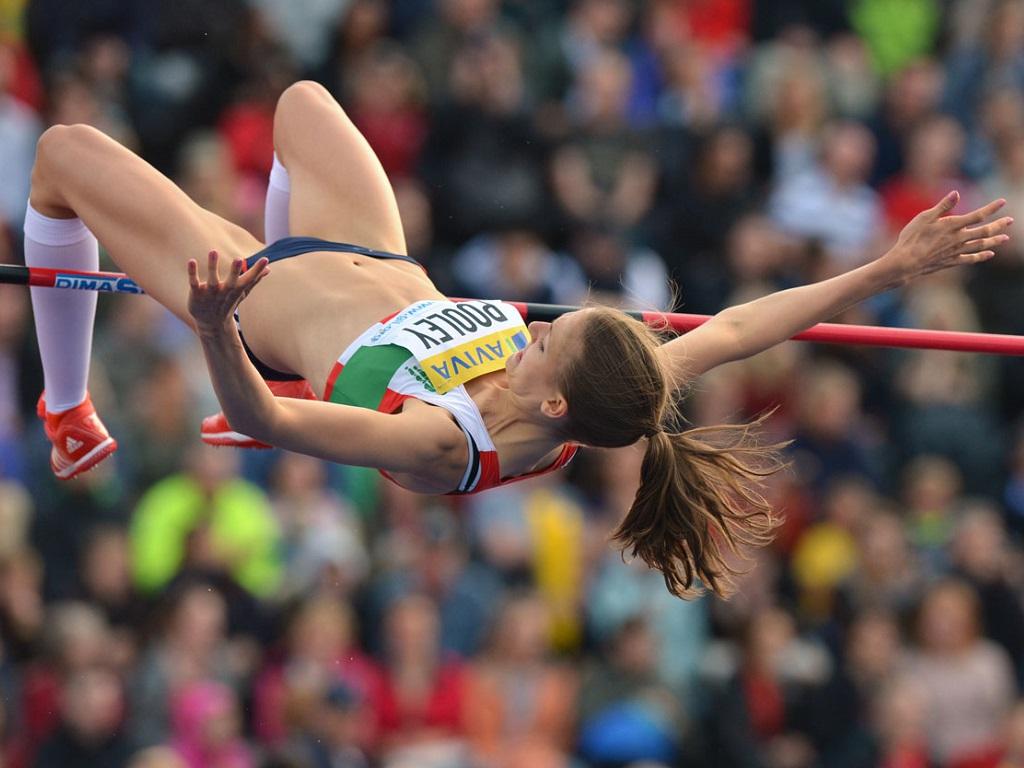}\vspace{4pt}
			\includegraphics[width=2.85cm]{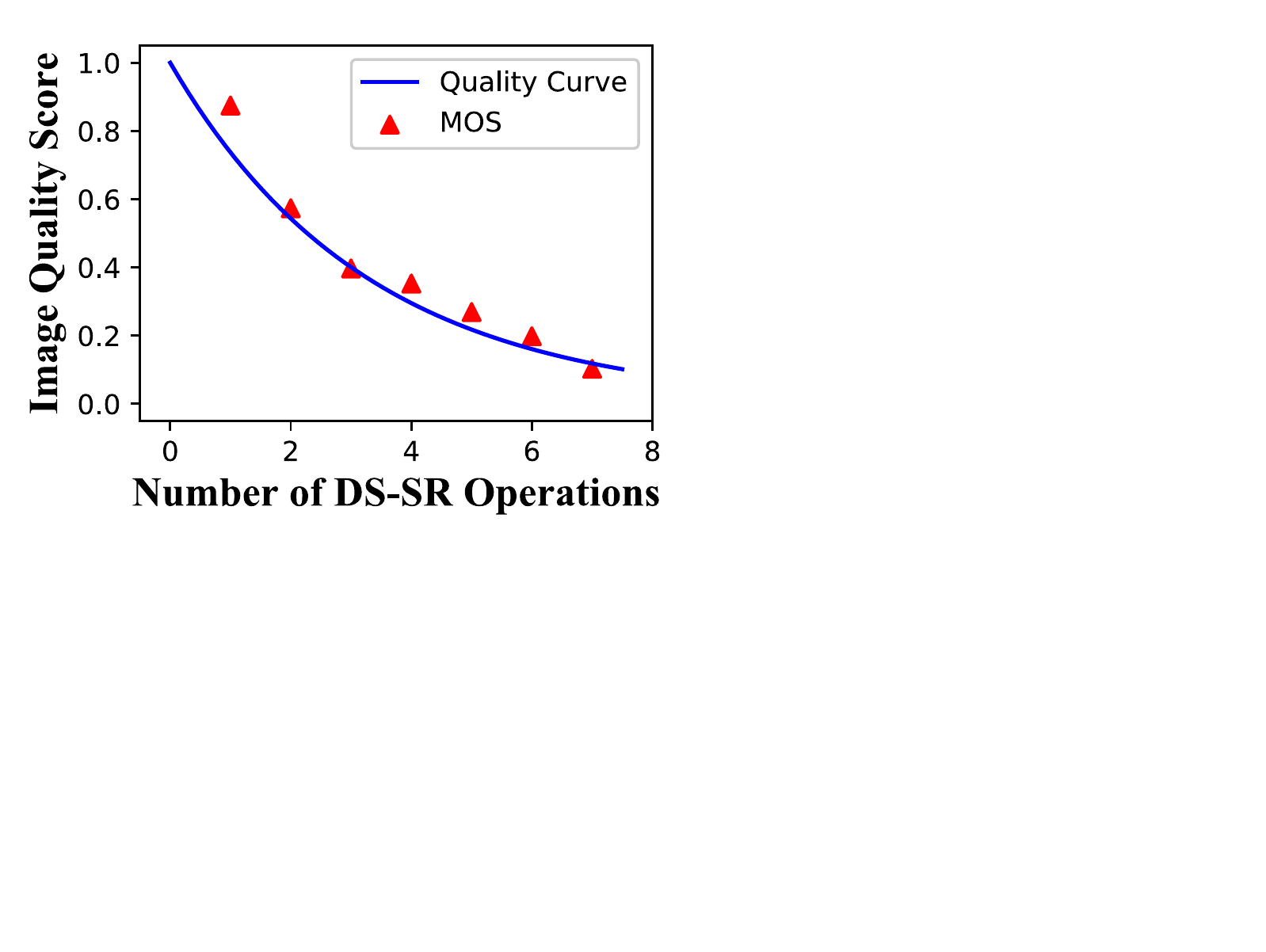}
	\end{minipage}}\hspace{-7.5mm}
	\subfigure[PLANTS: 0.9836]{
		\begin{minipage}{0.19\linewidth}
			\includegraphics[width=2.8cm]{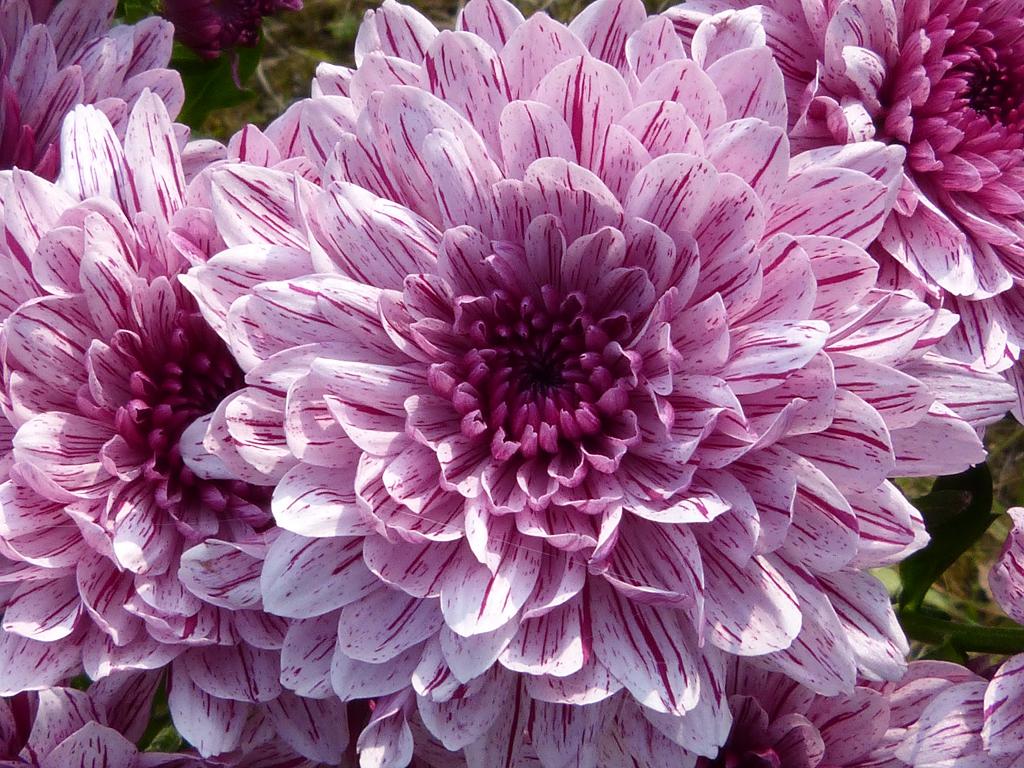}\vspace{4pt}
			\includegraphics[width=2.85cm]{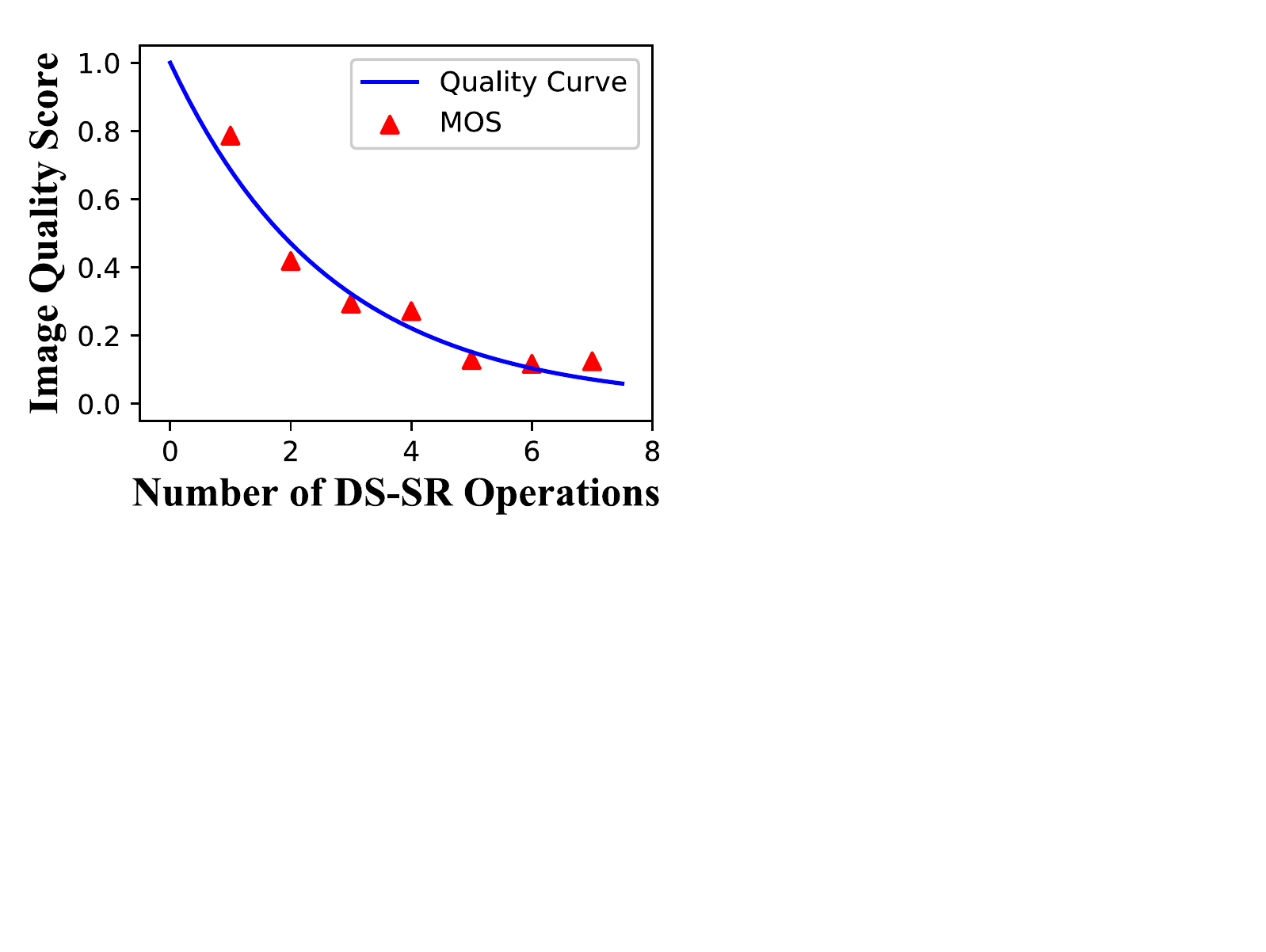}
	\end{minipage}}\hspace{-7.5mm}
	\subfigure[SCENERY: 0.9804]{
		\begin{minipage}{0.19\linewidth}
			\includegraphics[width=2.8cm]{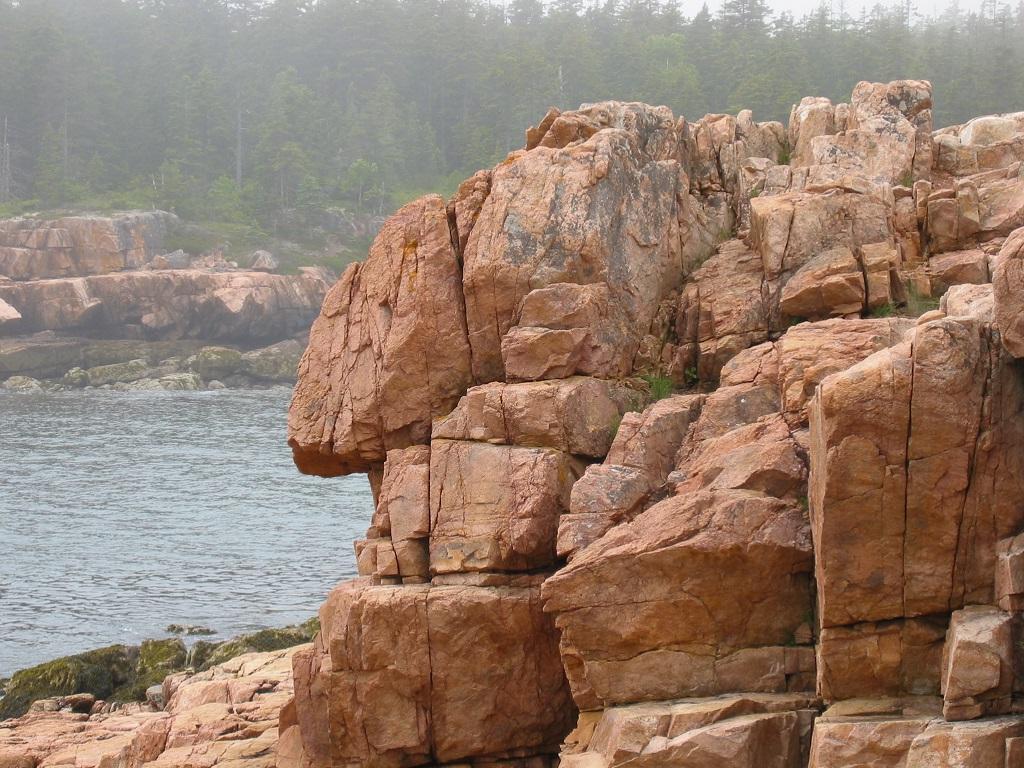}\vspace{4pt}
			\includegraphics[width=2.85cm]{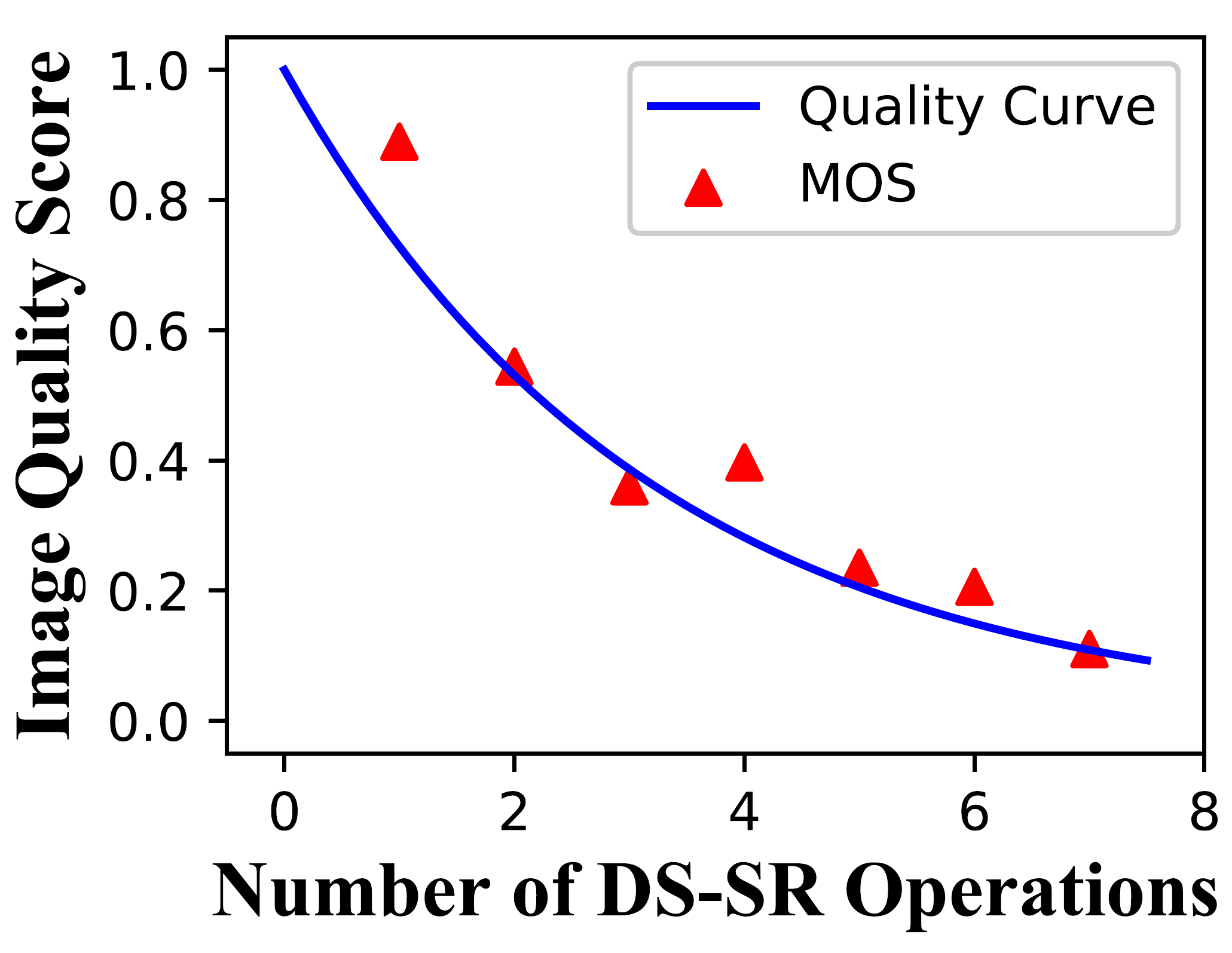}
	\end{minipage}}\hspace{-7.5mm}
	\centering
	\caption{Sample images and the corresponding MOS versus DS-SR iteration testing points with exponential decaying fitting curves.}
	\label{fig:curves}
\end{figure*}

\section{RELATED WORK}
%Although the Mean Square Error (MSE) is commonly used as a popular quality index, it still behaves as a poor criterion in many cases \cite{Yang20191}. 

Image SR has been an active research topic in recent years. Existing single image SR algorithms may be divided into three categories: interpolation-based \cite{Hou1979,Liu2012,Romano2014}, reconstruction-based \cite{Zhang2012,Vardan2016,Ren2017} and learning-based \cite{Dong2016,Kim2016,Lai2019,Ahn2018,Tian2020,Zhang2020}. Interpolation-based methods are simple, efficient and of low computational cost but has limited restoration performance, because such methods often introduce severe artifacts. Built upon models of prior domain knowledge, reconstruction-based methods suppress artifacts better, but often have much higher computational complexity. Learning-based methods learn LR-to-HR image mapping from training data.

During the past decades, numerous perceptual IQA metrics have been proposed to predict the visual quality of images with full-reference, reduced-reference and no-reference. Among them, the scope of application of full-reference metric is limited due to its requirement of unimpaired source. Recently, researchers have paid more attention to no-reference IQA. Early no-reference methods were based on Natural Scene Statistic (NSS) \cite{NSS2011}. Common NSS features include wavelet coefficients \cite{Moorthy2012}, locally normalized illumination coefficients \cite{Mittal2012}. Learning-based methods have been growing steadily \cite{Xu2016,Shao2016,Gu2018}, which automatically learn the mapping between image features and the perceptual quality. 
%Xu \textit{et al.} proposed a blind IQA method based on High Order Statistics Aggregation (HOSA), in which features of training images were extracted to build a codebook, and the image quality was estimated by feature encoding \cite{Xu2016}.

Meanwhile, Image Sharpness Assessment (ISA) technique has emerged as an effective method for IQA \cite{HVS-MaxPool2019}. Hassen \textit{et al.} identified the Local Phase Coherence (LPC) of images computed in wavelet transform domain to assess image quality \cite{LPC-SI2013}. Bahrami \textit{et al.} proposed a fast no-reference ISA method based on the standard deviation of weighted Maximum Local Variation (MLV) distribution of images \cite{MLV2014}. Blanchet \textit{et al.} introduced an indicator of Global Phase Coherence (GPC), which decreases with blur, noise, and ringing \cite{GPC2015}.  Li \textit{et al.} proposed a no-reference SPArse Representation-based Image SHarpness index (SPARISH), which used an overcomplete dictionary learned from natural images to measure the extent of blur \cite{SPARISH2016}. Hosseini \textit{et al.} designed two no-reference ISA metrics, called Synthetic-MaxPol \cite{Synthetic-MaxPool2018} and HVS-MaxPol \cite{HVS-MaxPool2019}, which were based on MaxPol convolution kernals and MaxPol filter library, respectively.

The use of deep learning has been a strong trend in recent IQA algorithms. Kang \textit{et al.} designed a shallow CNN to learn features from contrast normalized image patches \cite{Kang2014}. Ma \textit{et al.} proposed a multi-task end-to-end learning framework for no-reference IQA \cite{Ma2018}. Yang \textit{et al.} proposed an end-to-end SGDNet for no-reference IQA, which introduced saliency information to facilitate quality prediction \cite{Yang2019}. Yan \textit{et al.} integrated the NSS features prediction to the deep learning-based no-reference IQA to improve the representation and generalization ability \cite{Yan2019}. 

However, most existing IQA methods are not suitable for SR images, since they are designed for images degraded by common distortions such as compression, white noise and blur. The distortions produced by SR are often mixed and more sophisticated. The study \cite{Wang2017} has shown that the popular no-reference IQA metrics are difficult to predict the perceptual quality of SR image based on an SR Image Database (SRID).

In a lab testing environment, the pristine reference HR image may be available when evaluating SR algorithms, for which objective full-reference metrics are directly applicable. It was shown that the Mean Squared Error (MSE) is a poor criterion in many cases \cite{Yang20191}. In  \cite{Thapa2016}, Thapa \textit{et al.} adopted PSNR and SSIM to compare the performance of several SR algorithms. In SR benchmark study  \cite{Yang2014}, Yang \textit{et al.} analyzed the effectiveness of six common full-reference IQA indicators, and also demonstrated that existing full-reference metrics fail to match the perceptual qualities of HR images well. Small-scale subjective tests have also been carried out. Reibman \textit{et al.} evaluated SR enhanced image quality through subjective tests, and pointed out that the full-reference IQA metrics cannot always capture visual quality of HR image \cite{Reibman2006}.

In most practical scenarios, the reference HR images are not available, thus reduced-reference and no-reference IQA algorithms are preferred. Yeganeh \textit{et al.} extracted three sets of NSS to assess the image SR performance with the available LR image as a reference, but the interpolation factors were strictly limited to integers \cite{Yeganeh2015}. Zhou \textit{et al.} proposed a Quality Assessment Database for Super-resolved images (QADS) and an IQA method considering the structural and textural components of images \cite{Zhou2019}. Chen \textit{et al.} presented a hybrid quality metric for non-integer image interpolation that combined both reduced-reference and no-reference philosophies \cite{Chen2018}. Fang \textit{et al.} introduced a reduced-reference quality assessment method for image SR by predicting the energy and texture similarity between LR and HR images \cite{Fang2016}. In \cite{Tang2019}, Tang \textit{et al.} proposed another reduced-reference IQA algorithm for SR reconstructed images with information gain and texture similarity combining saliency detection. Ma \textit{et al.} proposed a no-reference metric by supervised learning on a large set of reference-free HR images \cite{Ma2016}. In addtion, Fang \textit{et al.} designed a Blind model based on CNN for SR-IQA (BSRIQA) \cite{Fang2018}. 
%Recently, Zhou \textit{et al.} considered both the structural and textural characteristics of the SR images into the proposed network to predict the perceptual quality of SR images in an NR manner \cite{Zhou2020}. 

Despite the significant effort, existing SR-IQA methods are limited in three aspects. First, most of them are constrained to integer scaling factors. Second, the public image SR databases are limited in size, making it difficult to train SR-IQA models without encountering serious overfitting problems. Third, there is no deep learning based model for reduced-reference SR-IQA.

Focusing on the issues above, we firstly establish a large-scale image SR database, in which the scaling factor of HR images can be arbitrary. A reduced-reference CNN-based SR-IQA method is then proposed, which combines with LR images as references and is trained on the constructed large-scale database.

\section{SR Image Database Construction}

Learning-based IQA methods desire large-scale databases for training. The main challenge in building such databases is how to label a large number of images with quality ratings. Human subjective testing is desirable, but is time-consuming and expensive. Moreover, the fatigue effect when labeling large-scale datasets often affects the consistency and reliability of subjective ratings. Here we opt to a semi-automatic approach, aiming for largely reducing of workload of subjective testing.

\subsection{The Exponential Law in Image SR}

Presumably, when an image is going through a downsampling (DS) followed by SR interpolation process iteratively, the quality of the SR interpolated images decreases with iteration. To verify this effect, we apply the DS-SR operation iteratively on natural images of diverse content. We then conduct subjective tests to obtain Mean Opinion Score (MOS) values. Interestingly, we find that the MOS value decreases with iteration, approaximately following an exponentially decaying curve (with very few outliers), regardless of the image content or interpolation method, as shown in Fig. \ref{fig:curves}. Let $Q$ and $t$ denote the quality score and the number of SR iterations respectively, we have

%As a common knowledge, an interpolated HR image with larger interpolation factor tends to show lower visual quality. Inspired by this, we perform iterative downsample-and-SR (DS-SR) on natural images with diverse contents. Ideally, the visual quality should be degraded due to prolonged information loss. To measure the quality degradation, we conduct subjective tests to obtain Mean Opinion Score (MOS) values when the DS-SR operation is continuously performed. After normalization, the MOS values decrease exponentially as the iterations of DS-SR operation increase, which can be approximated as a negatively exponential curve. This fitting results in a high correlation to MOS for different types of images, as shown in Fig. \ref{fig:curves}. Few outliers hardly affect the conclusion. Let $Q$ and $t$ denote the quality score and the number of SR iterations respectively, the quality function is as follows:
\begin{equation}
  Q(t) = e^{-bt},  t\geq0,
  \label{eq:1}
\end{equation}
where $b$ is a positive constant depending on image content and SR method. We normalize the quality of pristine HR images to 1, thus all curves pass through a fixed point: $Q$ (0) = 1. Given $b$, the quality scores of a set of HR images can be quickly acquired. Therefore, this observation offers promises to significantly reduce the workload of image labeling. 

Inspired by the exponential decaying relationship described above, we carry out a dedicated experimemt to investigate the feasibility of using the relationship for semi-automatical labeling, which may greatly reduce the workload of subjective ratings. Firstly, we randomly select 300 images of 30 natural scenes and perform iterative DS-SR on these images. Secondly, we utilize two approaches to obtain the subjective scores. In the first approach, 21 subjects are asked to score a subset of HR images. Using the exponential decay law of Eq. (\ref{eq:1}), we interpolate the MOS values of the remaining images. Detailed information of this approach is illustrated in the following Section \uppercase\expandafter{\romannumeral3}. \textit{C}. In the second approach, all images are scored by all subjects to obtain all MOS values. Thirdly, we compare the MOS values obtained by the above two approaches: semi-automatic rating and full subjective test.

The results are evaluated using Pearson Linear Correlation Coefficient (PLCC), Spearman Rank-order Correlation Coefficient (SRCC) and Kendall Rank Correlation Coefficient (KRCC), as shown in Table \ref{tab:curve_corr}, which indicates an encouragingly high correlation between the two approaches. To provide a reference point, we also compute the correlations between the scores given by each individual subject and MOS. The average across all subjects, as reported in Table \ref{tab:curve_corr}, is a meaningful indicator of the performance of an average subject. Clearly, the correlation between the semi-automatic rating and MOS is higher than an average subject, another strong evidence that helps justify the semi-aotumatic rating approach. Given that the semi-aotumatic approach can greatly reduce the workload of human subject labelings, it offers great promise to generate quality labels with the absence of excessive subjective testing. Furthermore, the proposed DISQ model (details given later) trained on the SISAR database created from the semi-automatic rating approach well generalizes to produce superior performance on other databases, which again validates the usefulness of the proposed semi-automatic rating method.

%which implies the interchangeability of them. It is also noted that the semi-automatic rating approach greatly reduces the labeling workload of subjects. Therefore, the semi-automatic rating approach based on Eq. (\ref{eq:1}) is feasible to develop a quality database of image SR. In Table \ref{tab:curve_corr}, we also report the average correlation for subjects, which represents the average labelling accuracy in the full subjective test. Obviously, the accuracy of semi-automatic rating is superior to individual subject labelling, which also evidences that the effectiveness of semi-automatic rating. In addition, the proposed deep model trained with our database shows superior performance on other databases, which proves the effectiveness of our database from the side. In conclusion, the reliability of our large-scale SISAR database can be fully guaranteed.
 \begin{table}[h]
  \centering
  \caption{PLCC evaluation between Semi-Automatic Rating and Full Subjective Test}
  \label{tab:curve_corr}
 \setlength{\tabcolsep}{2mm}
\begin{tabular}{cccccc}
\toprule
Image Types	&LR Images	&HR Images	&PLCC	&SRCC	&KRCC \\ \midrule
Animals	&6	&60	&0.9776	&0.9716	&0.8667	\\
Buildings	&6	&60	&0.9925	&0.9731	&0.8747	\\
Humans	&6	&60	&0.9656	&0.9609	&0.8396	\\
Sports	&6	&60	&0.9842	&0.9725	&0.8776	\\
Plants	&3	&30	&0.9797	&0.9817	&0.9040	\\
Scenery	&3	&30	&0.9731	&0.9791	&0.8915	\\ \midrule
\multicolumn{3}{c}{Average}	&0.9825	&0.9719	&0.8619	\\ \midrule
\multicolumn{3}{c}{Average Subject Performance}	&0.9510	&0.9314	&0.8184	\\
\bottomrule
\end{tabular}
\end{table}

\subsection{Source Images and SR Algorithms}

The first step to construct the database is to collect a batch of HR images generated by SR algorithms. In the semi-automatic rating approach, it is achieved by iterative DS-SR on source images. The source images are selected to cover diversified scenes including animals, buildings, humans, sports, plants and scenery. In total, 100 natural images of $1024\times 768$ resolution are selected. Then, four typical SR algorithms are employed to generate HR images, including two interpolation-based methods (BICUBIC \cite{Hou1979}, RLLR \cite{Liu2012}) and two learning-based methods (SRCNN \cite{Dong2016}, VDSR \cite{Kim2016}). In particular, SRCNN and VDSR are constrained to image SR with integer scaling factors. We integrate these two methods and the BICUBIC algorithm to achieve non-integer image SR. In order to obtain HR images with varying qualities, three scaling factors,  1.5, 2 and 2.7, are utilized in the iterative DS-SR process. Hence, the scaling factor 2 is supported by BICUBIC, RLLR, SRCNN and VDSR while the other two factors are supported by BICUBIC, RLLR, SRCNN+BICUBIC and VDSR+BICUBIC, as shown in the Table \ref{tab:database}.

\begin{table}[t]
 \centering
  \caption{Composition of SISAR Database}
  \label{tab:database}
  \setlength{\tabcolsep}{1mm}
\begin{tabular}{ccccc}
\toprule
Source Images  & Algorithms   & Factors      & \begin{tabular}[c]{@{}c@{}}HR Images\\in a Batch\end{tabular}  & \begin{tabular}[c]{@{}c@{}}Total of\\HR Images\end{tabular} \\ \midrule
100             & \begin{tabular}[c]{@{}c@{}}BICUBIC\\RLLR\end{tabular}            & 1.5, 2, 2.7 & 8, 7, 6             & 4200 \\ \midrule
100             & \begin{tabular}[c]{@{}c@{}}SRCNN\\ VDSR\end{tabular}                 & 2           & 7                   & 1400       \\ \midrule
100             & \begin{tabular}[c]{@{}c@{}}SRCNN+BICUBIC\\ VDSR+BICUBIC\end{tabular} & 1.5, 2.7    & 8, 6                & 2800\\ \bottomrule
\end{tabular}
\end{table}

\begin{figure}[t]
  \centering
    \subfigure[MOS: 0.7769]{%0.8516
  \includegraphics[width=2.5cm]{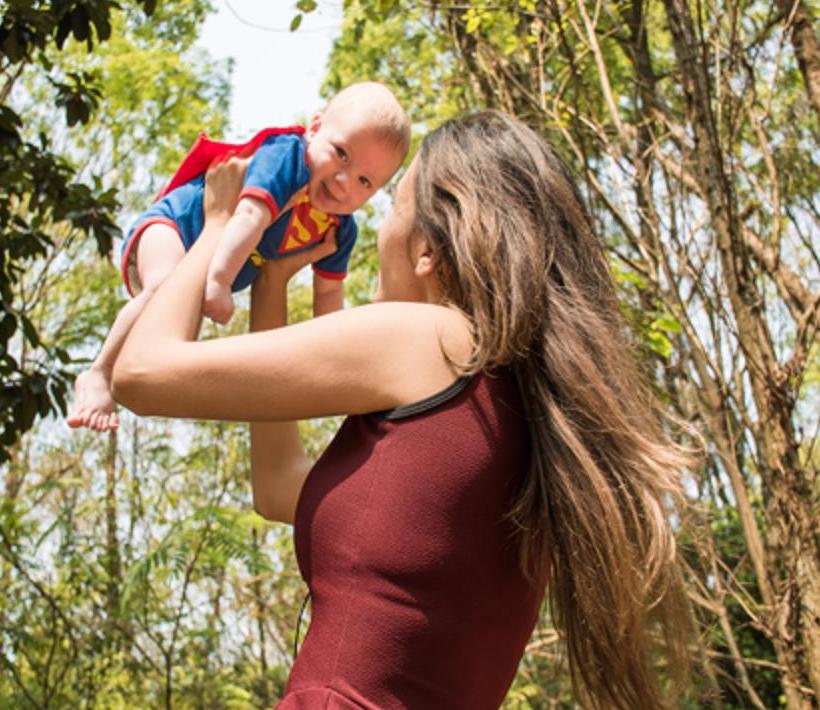}%building12/8316.jpg}
  }\vspace{-0.3mm}
      \subfigure[MOS: 0.6036]{% 0.7386
  \includegraphics[width=2.5cm]{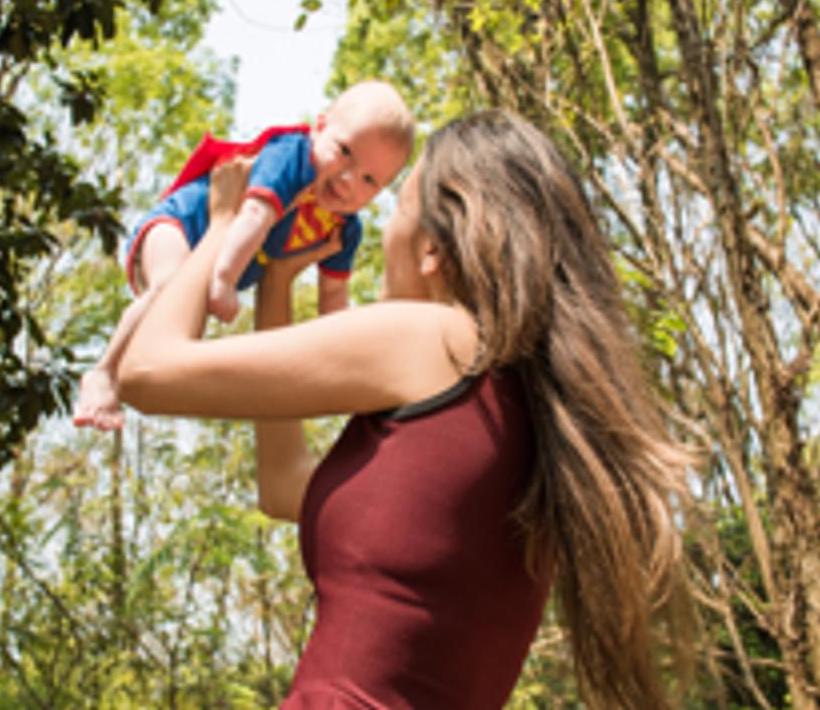}%building12/7186.jpg}
  }\vspace{-0.3mm}
    \subfigure[MOS: 0.4290]{%0.5845
  \includegraphics[width=2.5cm]{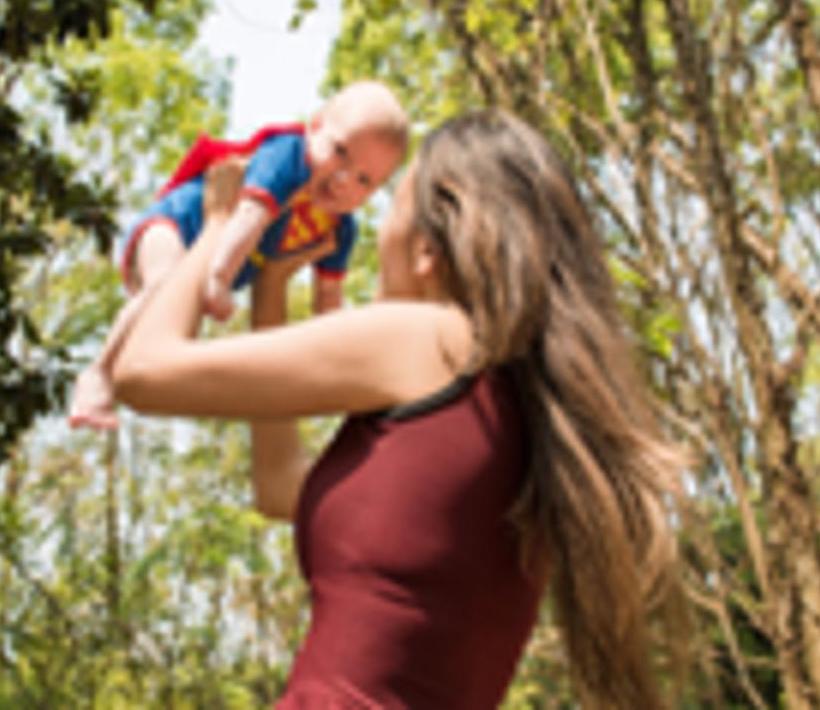}%building12/5845.jpg}
  }\vspace{-0.3mm}
    \subfigure[MOS: 0.2831]{
  \includegraphics[width=2.5cm]{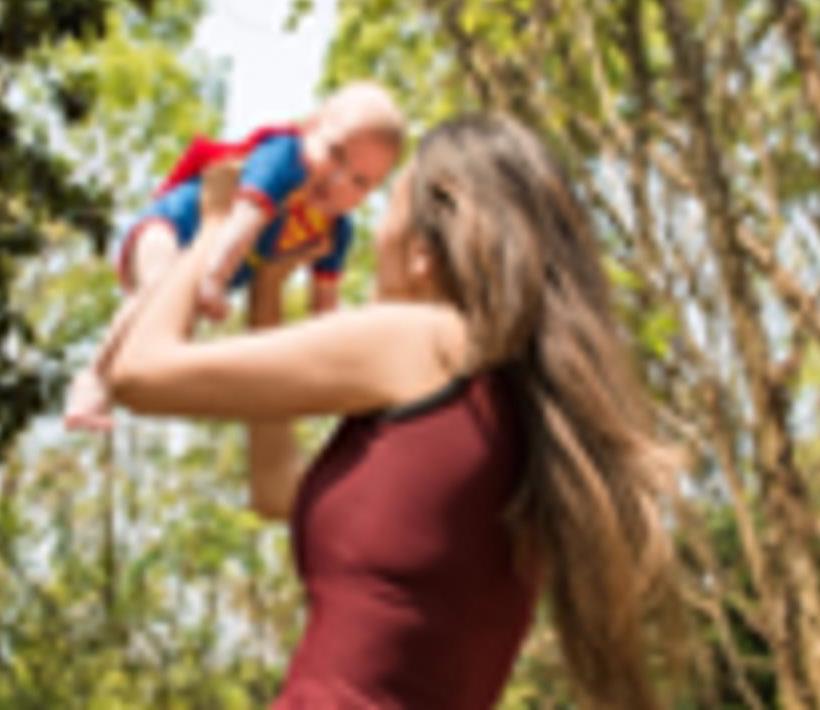}%building12/4338.jpg}
  }\hspace{-1mm}
      \subfigure[MOS: 0.1709]{
  \includegraphics[width=2.5cm]{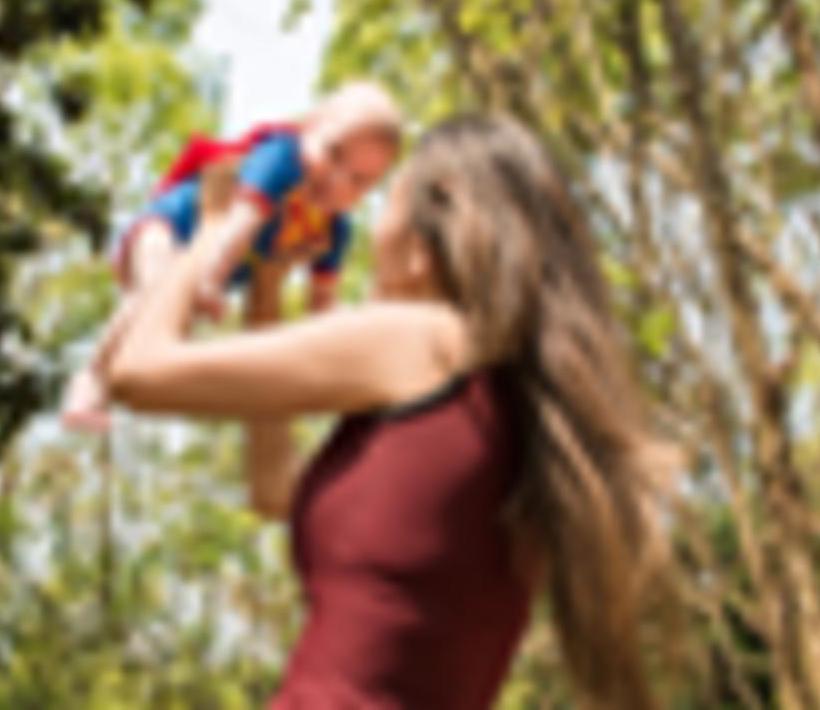}%building12/3520.jpg}
  }\hspace{-1mm}
    \subfigure[MOS: 0.0958]{
  \includegraphics[width=2.5cm]{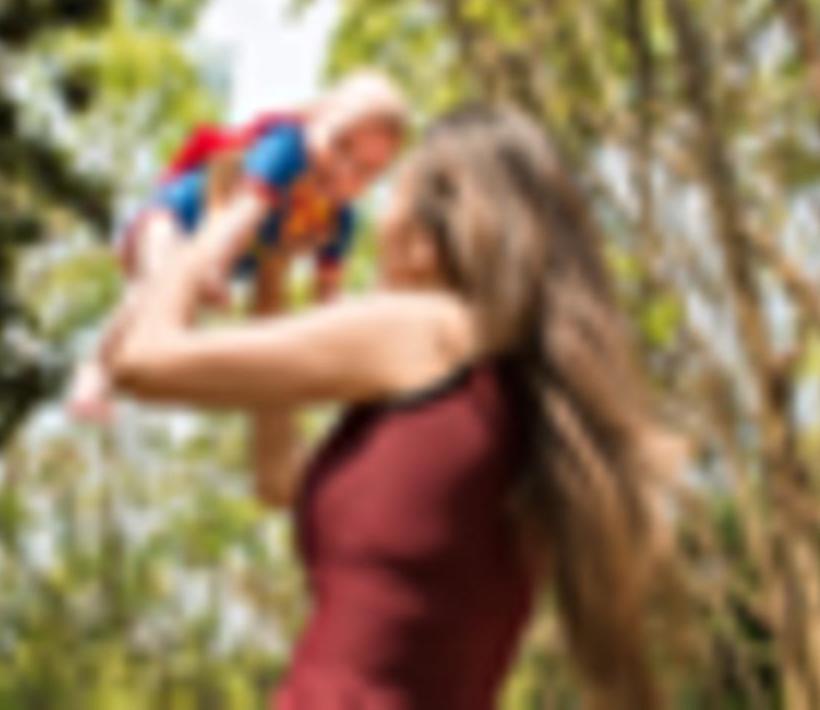}
  }
  \caption{Sample HR Images created by SR algorithms.}
  \label{fig:example}
\end{figure}

Our experiment as demonstrated in Fig. \ref{fig:curves} suggests that the image quality drops to an unacceptable level after certain number of DS-SR iterations. To avoid this, we set the maximum DS-SR iterations to 8, 7, 6 for the scaling factors 1.5, 2, 2.7, respectively. As a result, there are 8,400 HR images generated by image SR. An example of HR images of the same source is presented in Fig. \ref{fig:example}, where the SR algorithm BICUBIC is used with a scaling factor of 2.

\subsection{Image Quality Labeling}
We use the semi-automatic rating approach to label the quality scores of images in the SISAR database. According to Eq. (\ref{eq:1}), these scores can be calculated for given parameter $b$, which is dependent on image content and SR algorithm, and can be estimated through subjective test.

%with the parameter $b$ known for each batch of HR images. Let the normalized quality of source image be a maximum score 1, we can calculate the value $b$ with another point on the curve except $(0,1)$. This can be achieved by a subjective test.

For each (source image, SR algorithm, scaling factor) combination, we extract one HR test image for subjective testing, 25 subjects, aged between 20 and 30, are asked to give scores between 0 and 10 to assess the image quality. The HR images for test are randomly shown on the same screen together with their corresponding original images, which are considered as  pristine references with quality score of 10. The experiment procedure follows the Double Stimulus Continous Quality Scale (DSCQS) method defined in ITU R BT. 500-13 \cite{Recommendation2012}. After statistical analysis, 2 participants are identified to be outliers and the remaining 23 subjective scores are averaged to an interpreted Mean Opinion Score (iMOS) for each test image. We normalize the iMOS values to [0,1] to obtain a point ($k$, iMOS) on the corresponding exponential decaying curve, where $k$ is the number of SR iterations. Substituting the point into Eq. (\ref{eq:1}), we obtain the parameter $b$. 

With the parameter $b$ and Eq. (\ref{eq:1}), we compute the iMOS values of all other HR images in the group. Eventually, this semi-automatic labeling process saves the workload of human subjective testing by a factor of approximately 7.

%The histogram of the iMOS values are shown in Fig. \ref{fig:histogram}. Therefore, all HR images can be labelled with about 1/7 human effort compared with full subjective test. Meanwhile, the accuracy of SISAR database is also guaranteed, as analyzed in Section \uppercase\expandafter{\romannumeral3}. \textit{A}.
\begin{figure}[t]
  \centering
  \includegraphics[width=8cm]{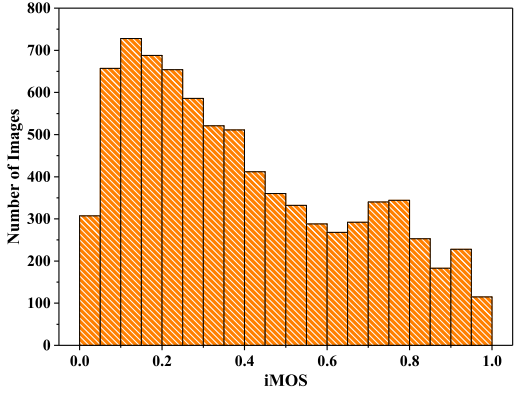}
  \caption{Histogram of the iMOS Values of the SISAR database.}
  \label{fig:histogram}
\end{figure}

\subsection{Summary of SISAR Database}

In total, the SISAR database contains 8,400 HR images, which are generated by three scaling factors with six types of SR algorithms that includes four SR algorithms (BICUBIC, RLLR, SRCNN and VDSR) and two combined SR operations (SRCCN+BICUBIC and VDSR+BICUBIC). The proposed database covers diverse image contents as shown in Table \ref{tab:img_num}, and the range of interpolated image resolution is from 864$\times$608 to 1054$\times$811. The histogram of the final iMOS values obtained by the proposed semi-automatic rating approach are shown in Fig. \ref{fig:histogram}, where it appears that the test HR images well cover the full range of quality levels.
\begin{table}[t]
 \centering
  \caption{Composition of SISAR database}
  \label{tab:img_num}
  \setlength{\tabcolsep}{1.2mm}
\begin{tabular}{@{}ccccccc@{}}
\toprule
 & Animals	& Buildings	 & Humans	 & Sports & Plants & Scenery \\  \midrule
Source Images & 20&20&20&20	&11 &9 \\
HR Images & 1680&1680&1680&1680	&924 &756 \\ \bottomrule    
\end{tabular}
\end{table}

\section{SR Image Quality Assessment Model}

\begin{figure*}[t]
  \centering
  \includegraphics[width=\textwidth]{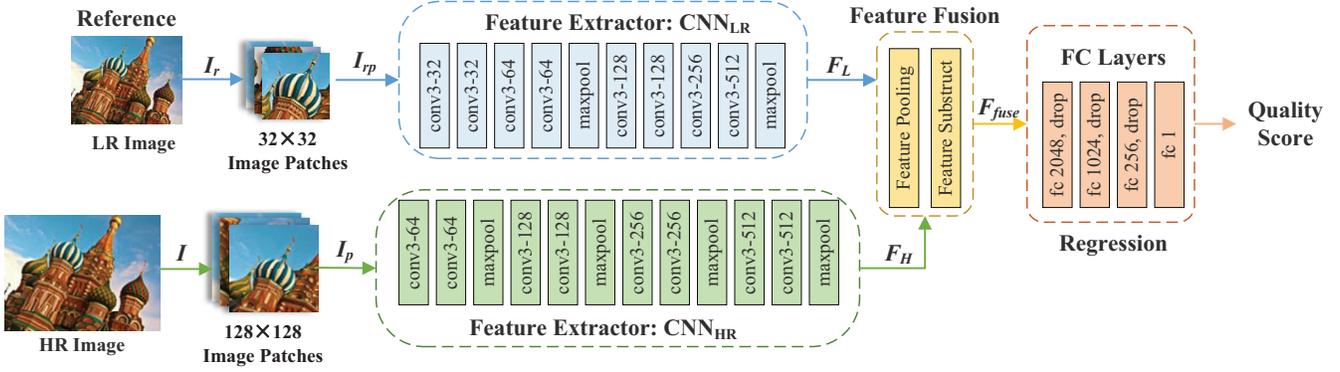}%scale=0.95
  \caption{Architecture of DISQ Framework.}
  \label{fig:framework}
\end{figure*}

We propose an end-to-end reduced-reference method for SR-IQA. The proposed model, namely Deep Image Super-resolution Quality (DISQ) index, is a two-stream CNN to evaluate the perceptual quality of HR images with LR images as references. The network architecture is illustrated in Fig. \ref{fig:framework}. Firstly, an HR image and its corresponding LR image are fed into each stream CNN to extracte image features. Then, a feature fusion step is conducted in order to combine discriminative features for the quality regression. Finally, the fused features are regressed onto the quality score of HR image by fully connected layers. In the following subsections, the network structure, feature fusion and model training will be discussed in detail.

\subsection{Network Architecture}

%\begin{figure*}[t]
 %\centering
  %\includegraphics[scale=0.8]{fig/network256.jpg} %[width=\textwidth]
  %\caption{Architecture of CNNs}
  %\label{fig:network}
%\end{figure*}

The proposed DISQ is built on an end-to-end learning framework that estimates the quality score of a test HR image with the HR image and LR image as inputs. The role of LR image is to provide reference information that may help the quality assessment of HR image. Considering the different scaling factors, we first split them into small non-overlapping image patches and input to the feature extractors to generate feature maps representing the whole images. Inspired by the architecture of VGG-16 network \cite{Simonyan2014}, we design our CNN for SR-IQA with a similar philosophy. %in Fig. \ref{fig:network}.

The proposed network contains two convolutional modules, a feature fusion module and a fully connected module. The convolutional module is utilized as feature extractor to learn various features from inputs, and the fully connected module is used to derive global features to regress quality scores. The feature fusion step is the bridge to connect these modules.

In the convolutional modules, all convolutions have a kernel size of 3$\times$3 with altered kernels. Combining with max-pooling layers with a stride of 2, the network extracts local features while reducing the size of the feature maps. We employ Rectified Linear Unit (ReLU) as the nonlinear activation function. The first convolutional module, named as ${\rm CNN}_{\rm LR}$, is designed to learn features of LR images. The inputs are the patches of an LR image with the size of 32$\times$32. Another convolutional module, denoted as ${\rm CNN}_{\rm HR}$, is a feature extractor of HR images. We divide each HR image into 128$\times$128 patches as inputs to generate feature maps of a whole HR image. To facilitate feature fusion, the two CNNs contain 2 and 4 max-pooling layers correspondingly, and several convolutional layers to obtain features with similar shapes.

After feature fusion, the fused features are fed into the fully connected module to be regressed onto the perceptual scores. There are four fully connected layers with 2048, 1024, 256 and 1 neurons, respectively. We apply dropout into the first three fully connected layers with a probability of 0.5. By randomly masking out the neurons, dropout helps prevent overfitting. The last layer is a simple linear regression with a scalar output that predicts the quality score.

\subsection{Feature Fusion}

We adopt \begin{math} {\rm CNN}_{\rm LR}\end{math} and \begin{math} {\rm CNN}_{\rm HR}\end{math} as the feature extractors to produce the feature maps \begin{math} F_{\rm H} \end{math} and \begin{math} F_{\rm L} \end{math} directly from the input HR image patches \begin{math}  I_{\rm p} \end{math} and LR image patches \begin{math}  I_{\rm rp} \end{math}, respectively. According to the internal structure of two convolutional modules, we can calculate the shape of output feature maps.
\begin{equation}
\begin{split}
\textbf{Feature } &\textbf{Extraction:} \\
 F_{\rm L} &= {\rm CNN}_{\rm LR}(I_{\rm rp}; \theta_1), {\rm shape} = (N_{\rm L}, 8, 8, 512), \\
 F_{\rm H} &= {\rm CNN}_{\rm HR}(I_{\rm p}; \theta_2), {\rm shape} = (N_{\rm H}, 8, 8, 512).
    \end{split}
\end{equation}
Here $\theta_1$ and $\theta_2$ indicate the parameters of CNNs. $N_{\rm H}$ and $N_{\rm L}$ denote the numbers of HR and LR image patches. The $F_{\rm H}$ and $F_{\rm L}$ are collections of image patch features. 

In order to transform the patch-level features into the image-level features, we first perform a pooling operation before image feature fusion. Specifically, the feature map is pooled into one mean tensor, max tensor and min tensor in the first dimension. Then the three tensors are concatenated into a new feature set without any further modifications.
\begin{equation}
\begin{split}
&\textbf{Feature } \textbf{Pooling:} \\
&F_{\rm mean} = {\rm mean}(F; {\rm axis}=1), {\rm shape} = (1, 8, 8, 512), \\
&F_{\rm max} = {\rm max}(F; {\rm axis}=1), {\rm shape} = (1, 8, 8, 512),\\
&F_{\rm min} = {\rm min}(F; {\rm axis}=1), {\rm shape} = (1, 8, 8, 512),\\
&F_{\rm pool} = (F_{\rm mean}, F_{\rm max}, F_{\rm min}), {\rm shape} = (3, 8, 8, 512).
    \end{split}
\end{equation}

After feature pooling, there are two pooled features $F_{\rm Hpool}$ and $F_{\rm Lpool}$ with identical structure created by $F_{\rm H}$ and $F_{\rm L}$, respectively. We fuse the extracted image feature maps $F_{\rm Hpool}$ and $F_{\rm Lpool}$ before inputting to the regression part of the network. In image SR, the HR images are reconstructed based on the global information of LR images, which renders the difference between HR and LR image features to a meaningful representation in the feature space. In order to calculate their distance, the feature fusion step follows  \cite{Bosse2018}:
\begin{equation}
  \textbf{Feature Fusion: } \quad F_{\rm fuse} = F_{\rm Hpool} - F_{\rm Lpool}.
  \label{eq:4}
\end{equation}

\begin{figure}[h]
  \centering
  \includegraphics[width=9.2cm]{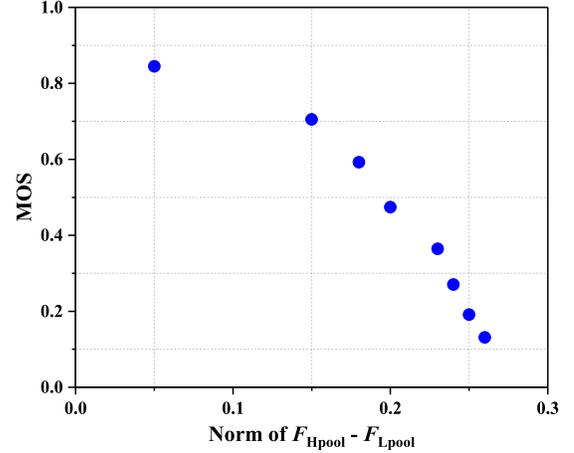}%scale=0.95
  \caption{The Correlation between MOS and the Norm of $ F_{\rm Hpool}-F_{\rm Lpool}$.}
  \label{fig:norm}
\end{figure}

Idealy, $F_{\rm Lpool}$ in Eq. (\ref{eq:4}) should be replaced by $F_{\rm Hpool}^0$, where $F_{\rm Hpool}^0$ denotes the pooled feature set of the perfect HR image. However, this perfect image is unfortunately unavailable. To investigate the impact of replacing $F_{\rm Hpool}^0$ with $F_{\rm Lpool}$, we carry out two empirical studies.
%in typical image SR problem. Therefore, we have to replace the $ F_{\rm Hpool}^0$  by $ F_{\rm Lpool}$. This operation is plausible because: (1) The $ F_{\rm Hpool}^0$  and $ F_{\rm Lpool}$ are sufficiently close in feature space; (2) The difference between $ F_{\rm Hpool}$ and $ F_{\rm Lpool}$ is correlated to subjective ratings. The two statements can be validated with the Table \ref{tab:distance} and Fig. \ref{fig:norm}, respectively. In Table \ref{tab:distance}, 
First, we perform test on a reference image and several corresponding LR images to calculate the difference of $F_{\rm Hpool}^0$ and $F_{\rm Lpool}$. The results are given in Table \ref{tab:distance}, which shows that the MSE of the two features are negligible relative to $F_{\rm Hpool}-F_{\rm Lpool}$.
%On the other hand, Fig. \ref{fig:norm} shows a high 
Second, we compute the correlation between the norm of $ F_{\rm Hpool}-F_{\rm Lpool}$ and subjective scores of images. The results shown in Fig. \ref{fig:norm} suggest that the correlation is strong. These empirical studies lead us to adopt Eq. (\ref{eq:4}). Furthermore, the effectiveness %Therefore, Eq. (\ref{eq:4}) can be utilized as a feature fusion method in our model. In addition, the efficiency 
of this fusion method is also validated by experiments in the following Section \uppercase\expandafter{\romannumeral5}. \textit{C}.

\begin{table}[h]
 \centering
  \caption{MSE between $F_{\rm Hpool}^0$ and  $F_{\rm Lpool}$, which is relatively small compare with the MSE between $F_{\rm Hpool}$ and  $F_{\rm Lpool}$ (~$10^{-2}$)}
  \label{tab:distance}
  \setlength{\tabcolsep}{2.2mm}
\begin{tabular}{@{}ccccc@{}}
\toprule
                    & \begin{tabular}[c]{@{}c@{}}Downsample\\by 2\end{tabular} & \begin{tabular}[c]{@{}c@{}}Downsample\\by 2.5\end{tabular} & \begin{tabular}[c]{@{}c@{}}Downsample\\by 2.7\end{tabular} & \begin{tabular}[c]{@{}c@{}}Downsample\\by 3\end{tabular} \\ \midrule
MSE& 3.357E-03                                                                               & 3.363E-03                                                                            & 3.401E-03                                                                         & 3.380E-03                                                                          \\ \bottomrule
\end{tabular}
\end{table}

%\begin{figure*}[h]
%	%\vspace{0.05cm}
%	\small
%\centering
%\subfigure[]{
%	\label{fig16:subfig:a} %% label for first subfigure
%	\includegraphics[height=8cm]{./fig/Database_Fig/SISAR.jpg}}\hspace{0.2cm}
%\subfigure[]{
%	\label{fig16:subfig:b} %% lael for second subfigure
%	\includegraphics[height=8cm]{./fig/Database_Fig/Waterloo.jpg}}\hspace{0.2cm}
%\subfigure[]{
%	\label{fig16:subfig:c} %% label for first subfigure
%	\includegraphics[height=8cm]{./fig/Database_Fig/CVIU.jpg}}\hspace{0.2cm}
%\subfigure[]{
%	\label{fig16:subfig:b} %% lael for second subfigure
%	\includegraphics[height=8cm]{./fig/Database_Fig/QADS.jpg}}\hspace{0.2cm}
%\vspace{0.1cm}
%\centering
%	\caption{\small Typical HR images in (a) SISAR, (b) Waterloo-15, (c) CVIU-17 and (d) QADS.}
%	\label{fig:ref}
%	
%	\vspace{0.15cm}
%\end{figure*}

\begin{figure*}[h]
	%\vspace{0.05cm}
	\small
\centering
\includegraphics[width=18cm]{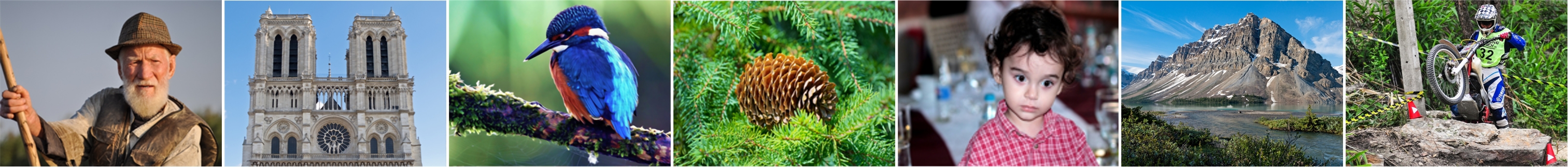}\vspace{0.15cm}
\includegraphics[width=18cm]{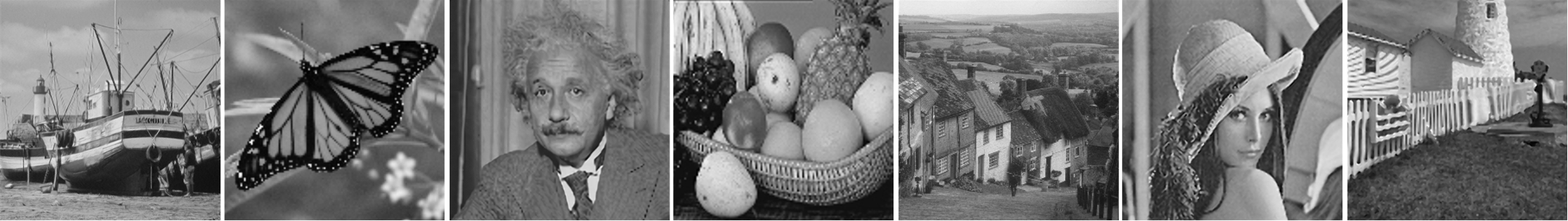}\vspace{0.15cm}
\includegraphics[width=18cm]{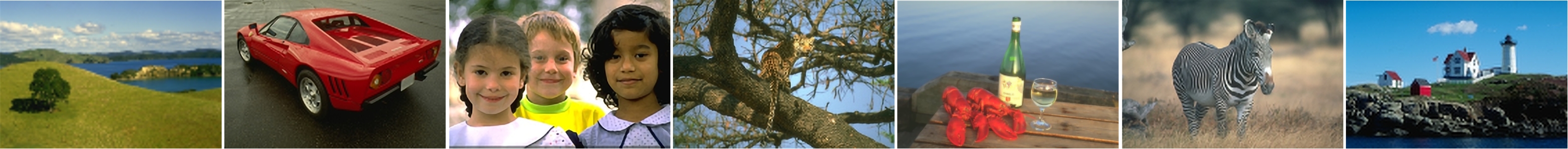}\vspace{0.15cm}
\includegraphics[width=18cm]{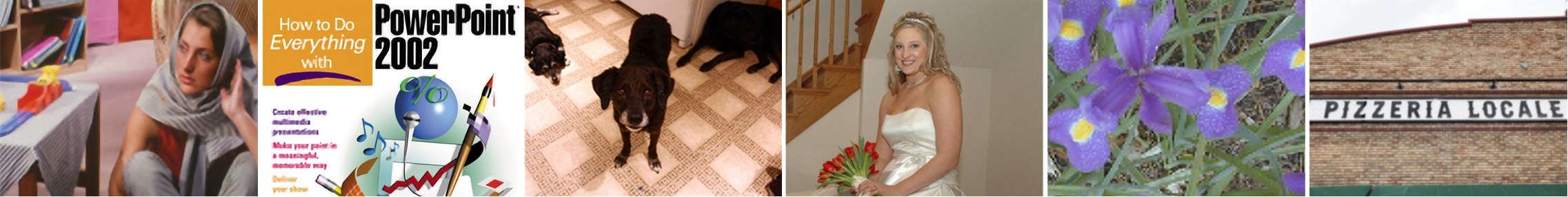}\vspace{0.15cm}
\centering
	\caption{\small Typical HR images of SISAR (line 1), Waterloo-15 (line 2), CVIU-17 (line 3) and QADS (line 4).}
	\label{fig:ref}
	
	\vspace{0.15cm}
\end{figure*}

\subsection{Model Learning}

For an input HR image $I$ and a reference LR image $I_{\rm r}$, the proposed SR-IQA network $M$ is used to predict the perceptual quality of HR image $Q_{\rm pred}$:
\begin{equation}
  Q_{\rm pred} = M((I_{\rm r}, I); \theta),
\end{equation}
where $\theta$ indicates all parameters of this network.

Denote the ground truth quality of the input as $Q_{\rm gt}$. The training goal of network $M$ is to find the optimal parameter setting, so as to minimize the overall quality prediction loss between $Q_{\rm pred}$ and $Q_{\rm gt}$ of all HR images in the training dataset. We apply the MSE as loss function in the training process, which is widely used in various regression tasks. Moreover, in order to avoid overfitting, $l$2-norm is added to the loss as a penalty term. Thus, the loss function is:
\begin{equation}
  {\rm Loss} = \frac{1}{N} \sum_{i=0}^{N} \left\| Q_{\rm pred,i} - Q_{\rm gt,i} \right\| ^2 + \lambda \left\| \theta \right\| ^2,
\end{equation}
where the subscript $i$ of $Q_{\rm pred, i}$ and $Q_{\rm gt, i}$ represent the predicted quality and group truth of the $i$-th image, respectively. $\lambda$ is the regularization coefficient set to 0.0005 in our work.

The Adam optimizer \cite{2014Adam} is adopted to minimize the loss function. The main advantage of Adam is that after bias correction, the learning rate of each epoch has a definite range, which makes the parameters more stable. The learning rate is $\eta$=0.0001, and other parameters of Adam optimizer are of default settings. 
%: $\beta_1$=0.9, $\beta_2$=0.999. 

\section{EXPERIMENTAL RESULTS}

In this section, we evaluate the performance of the proposed DISQ model and compare it with 16 other IQA metrics on four datasets. The cross-database test is also performed to assess the generalizibility of the algorithms.

\subsection{Experimental Setups}

\textbf{Databases}: We train the DISQ model on the proposed SISAR database. 
To verify the generalizibility of the proposed method, we also employ three publicly available databases, Waterloo-15 \cite{Yeganeh2015}, CVIU-17 \cite{Ma2016} and QADS \cite{Zhou2019}, for cross-database validations. Typical images of these databases are listed in Fig. \ref{fig:ref}.

\textbf{SISAR} contains 8,400 HR images. We divide it into training and test sets with a 80/20 split and no content overlapped.

\textbf{Waterloo-15} is an interpolation image database containing 312 interpolated HR images and the corresponding LR images from 13 source images \cite{Yeganeh2015}. The HR images were created by eight interpolation algorithms combined with scaling factors of 2, 4 and 8.

\textbf{CVIU-17} is a collection of 180 LR images and 1,620 HR images, which were generated by nine SR methods and six integer scaling factors. There is another available SR image database, ECCV-14 \cite{Yang2014}, which contains 540 HR images. It should be noted that ECCV-14 is a subset of CVIU-17. Therefore, we only provide the experimental result on CVIU-17.

\textbf{QADS} is a super-resolved image database, which created 980 HR images using 21 image SR methods from 20 reference images.

The score ranges and types are not unified in these database, we choose the settings of SISAR as our standard in these experiments. Subjective scores on the other three databases are linearly scaled to the range of [0,1].

\textbf{Evaluation}: We use three common measurements to evaluate the performance of our algorithm by calculating the correlation between the subjective and objective quality scores: PLCC, SRCC and KRCC. PLCC is used to measure the accuracy of IQA algorithms. SRCC and KRCC are used to evaluate the monotonicity of quality predictions. For these metrics, a higher value up to 1 indicates better performance of a specific IQA method. The results in this work are the average values obtained by calculating the correlation coefficients based on different image contents.

\subsection{Performance Comparison}

\begin{table*}[t]
\centering
\caption{Performance Comparison of IQA Methods: The datasets used for training the test algorithms are denoted by ‘—’.}
\label{tab:comparison}
\begin{tabular}{@{}lccc|ccc|ccc|ccc@{}}
\toprule
	& \multicolumn{3}{c}{Waterloo-15}	& \multicolumn{3}{c}{CVIU-17}	&\multicolumn{3}{c}{QADS}	&\multicolumn{3}{c}{SISAR}         \\ \cmidrule(l){2-13} 
% & \multirow{-2}{*}{ \begin{tabular}[c]{@{}c@{}}SR-IQA\\(Y/N)\end{tabular}} 
\multirow{-2}{*}{Methods} & PLCC & SRCC & KRCC     & PLCC & SRCC & KRCC		& PLCC & SRCC & KRCC    & PLCC & SRCC & KRCC            \\ \midrule
DIIVIVE	& 0.6202 & 0.5827 & 0.4225		& 0.4602 & 0.4580 & 0.3310		&0.4404 &0.4654 &0.3184		&0.3349 &0.3201 &0.2576     \\
BRISQUE & 0.7527 & 0.7626 & 0.5682  	& 0.5025 & 0.4644 & 0.3557      &0.5242 &0.5461 &0.3830     &0.3408 &0.3487 &0.2790     \\
CNN-IQA &{0.6659} &{0.6579} &{0.4856}		& {0.2933} & {0.2894} & {0.2141}		&{0.4037} &{0.3734} &{0.2694}        &{0.2345} &{0.2752}&{0.1910}	\\
HOSA    & 0.8287 & {\color[HTML]{CB0000} \textbf{0.8236}} & {\color[HTML]{CB0000} \textbf{0.6375}} 			& {\color[HTML]{3166FF} \textbf{0.5927}} & 0.5450 & 0.4169     &0.6343 &0.6409 &0.4598				&0.3299 &0.3180 &0.2629       \\
LPC-SI  & 0.8017 & 0.4689 & 0.4188		& 0.4547 & 0.4164 & 0.3294		&0.5027 &0.4902 &0.3358		&0.8562 &0.8076 &0.6267		\\
MLV		& 0.7313 & 0.3508 & 0.2903		& 0.5655 & 0.4628 & 0.3675		&0.2471 &0.2456 &0.1644		&0.7085 &0.6399 &0.5631		\\
GPC		& {\color[HTML]{3166FF} \textbf{0.8310}} & 0.5058 & 0.4289		&0.5878 &{\color[HTML]{3166FF} \textbf{0.5576}} &{\color[HTML]{3166FF} \textbf{0.4549}}		&0.4002 &0.4470 &0.3116		&0.7057 &0.8755 &0.7221		\\
SPARISH & 0.6988 & 0.6584 & 0.4924		& 0.4711 & 0.4390 & 0.3430		&0.5849 &0.6530 &0.4713		&0.8750 &0.8933 &{\color[HTML]{3166FF} \textbf{0.7373}}		\\
Synthetic-MaxPol &0.5909 &0.2938 &0.2249		&0.4920 &0.4537 &0.3536		&0.4057 &0.3845 &0.2645		&0.4607 &0.3305 &0.2867		\\
HVS-MaxPol &0.7905 &0.7272 &0.5401		&0.5448 &0.5557 &0.4343		&0.5918 &0.5876 &0.4098		&0.3880 &0.3151 &0.2788		\\
FQPath	&0.8038 &0.7544 &0.5672		&0.5122 &0.5111 &0.3884 	&0.4879 &0.4750 &0.3232 	&0.3200 &0.3429 &0.3127		\\
FocusLiteNN	 &{\color[HTML]{CB0000} \textbf{0.8368}} &0.5359 &0.4441 	&0.5338 &0.4464 &0.3361 	&0.3775 &0.3721 &0.2533		&{\color[HTML]{3166FF} \textbf{0.9123}} &{\color[HTML]{3166FF} \textbf{0.9016}} &0.7361		\\ 
NSS-SR  & {—} & —  & {—}		& {0.5504} & {0.4993} & {0.3616}		& {0.3670} & {0.2160} & {0.1424}      	& {0.3706} & {0.3280} & {0.2635}	\\
HYQM    & 0.4108 & 0.2096 & 0.2004		&0.2225 & 0.1125 & 0.0941		&0.4506 & 0.4443 & 0.2973		& 0.3660 & 0.2732 & 0.2112	\\ 
LNQM    & 0.7488 &0.7022  & 0.4974      & —     & —      & —   	&{\color[HTML]{3166FF} \textbf{0.7220}} & {\color[HTML]{3166FF} \textbf{0.7274}} & {\color[HTML]{3166FF} \textbf{0.5394}}		& 0.5192 &0.4944 &0.3449  \\
BSRIQA  &0.6991 &0.6736 &0.4689		&—      &—      &—       &0.5863 &0.5841 &0.4173		& 0.6906 &0.4341  &0.3099 \\ \midrule
DISQ    & 0.8232 & {\color[HTML]{3166FF} \textbf{0.7897}} & {\color[HTML]{3166FF} \textbf{0.5770}} 			& {\color[HTML]{CB0000} \textbf{0.6440}} & {\color[HTML]{CB0000} \textbf{0.7071}} & {\color[HTML]{CB0000} \textbf{0.5280}}  			&{\color[HTML]{CB0000} \textbf{0.8173}} &{\color[HTML]{CB0000} \textbf{0.7960}} &{\color[HTML]{CB0000} \textbf{0.6302}}			& {\color[HTML]{CB0000} \textbf{0.9143}} & {\color[HTML]{CB0000} \textbf{0.9124}} & {\color[HTML]{CB0000} \textbf{0.7742}}  \\ \bottomrule
\end{tabular}
\end{table*}

The performance evaluation results of the proposed DISQ model are listed in Table \ref{tab:comparison} and compared with other IQA methods, where the best and second-best results are shown in {\color[HTML]{CB0000} \textbf{red}} and {\color[HTML]{3166FF} \textbf{blue}}, respectively. The compared algorithms include four no-reference IQA methods (DIIVIVE \cite{Moorthy2012}, BRISQUE \cite{Mittal2012}, HOSA \cite{Xu2016}, CNN-IQA \cite{Kang2014}) designed for general distorted images, six no-reference ISA indicators (LPC-SI \cite{LPC-SI2013}, MLV \cite{MLV2014}, GPC \cite{GPC2015}, SPARISH \cite{SPARISH2016}, Synthetic-MaxPol \cite{Synthetic-MaxPool2018}, HVS-MaxPol \cite{HVS-MaxPool2019}), two focus quality assessment algorithms (FQPath \cite{FQPath2020}, FocusLiteNN \cite{wang2020focuslitenn}), and four related SR-IQA works (NSS-SR \cite{Yeganeh2015}, HYQM \cite{Chen2018}, LNQM \cite{Ma2016}, BSRIQA \cite{Fang2018}). 
%The compared metrics are all fully supervised learning-based methods.%
The source codes of these metrics are obtained from the authors’ public websites. There are two other SR-IQA metrics \cite{ Fang2016, Tang2019} without publicly available codes and parameters, thus they are not compared in the table. %It is noteworthy that compared methods have no full-reference IQA methods such as PSNR and SSIM, because most of them require the same sizes of the reference and distorted images, and the size of HR images in our database varies with SR iterations. 
From Table \ref{tab:comparison}, we have the following observations: 

First, FocusLiteNN obtains the highest PLCC on the Waterloo-15 database, and the HOSA achieves the best performance on SRCC and KRCC. In this database, all images were interpolated with integers, which implies that the pixels at LR image will be reused in HR image. An SR-IQA algorithm can utilize this feature to enhance their evaluation performance. However, the proposed DISQ algorithm is designed for general purpose SR without prior knowledge of scaling factors. Nevertheless, its performance is still close to the best. In particular, DISQ ranks the second in terms of SRCC and KRCC.

Second, DISQ achieves a good performance on the CVIU-17 database, while other metrics present moderate performances. CVIU-17 was created by several popular image SR algorithms, in which HR images are not limited to the common distortions. Therefore, the features modeled in common IQA methods cannot cover diverse visual contents. Although NSS-SR and NYQM are designed for image SR, they do not work well on this database, because the hand-crafted features of these methods are built more specifically for image interpolation.

Third, most non-SR-IQA methods are less effective on the QADS database. DISQ and LNQM, both learning-based, show excellent performance in comparison with other methods. Between them, DISQ utilizes the large-scale database and deep learning to effectively extract intrinsic features, and shows much better performance. The moderate performance of the CNN-based BSRIQA may be attributed to the small training database. %Compared with the training set of BSRIQA, ours contains richer image contents. In addition, it reveals that DISQ has a better generalization performance. 
 
Fourth, it can be clearly observed that DISQ outperforms other metrics on the SISAR database. FocusLiteNN achieves excellent performance, and most no-reference ISA algorithms show a good correlation with SISAR database. The reason is that the methods perform better on the local focused images in SISAR database than other metrics. However, their performance drops significantly on the other three databases. Among the SR-IQA methods, although LNQM and BSRIQA obtains better performance than NSS-SR and NYQM, they are not among the best performers. Both of them are trained on CVIU-17, in which the HR images were generated only by integer scaling factors. By contrast, SISAR contains HR images produced by non-integer scaling factors, which benefits the DISQ model that uses SISAR for training.

To further investigate model generalization ability, Table \ref{tab:average} reports the average performances of all algorithms under comparison on different databases. The performance of the algorithms above the dividing line are the average of four databases, and the rest are average of three, except for the database used for training the corresponding algorithm. As can be observed, the DISQ achieves the best average performance in cross-database experiments, and the PLCC and SRCC values are nearly 10\% higher than the second-best model. The results demonstrate the robustness of the proposed DISQ model.
\begin{table}[h]
\centering
\caption{Average Performance in Cross-Database Comparison}
\label{tab:average}
\setlength{\tabcolsep}{5mm}
\begin{tabular}{lccc}
\toprule
Methods 		& PLCC         		&SRCC			&KRCC    	\\  \midrule
DIIVIVE  	 	& 0.4639       		& 0.4566		&0.3324     \\
BRISQUE  		& 0.5301       		& 0.5305		&0.3965     \\
CNN-IQA  		& 0.3994          	& 0.4040		&0.2900     \\
HOSA     	 	& 0.5964			& 0.5819		&0.4443 	\\
LPC-SI			& 0.6538 			& 0.5458 		&0.4277		\\
MLV 			& 0.5631         	& 0.4248		&0.3326     \\
GPC 			& 0.6312         	& 0.5965		&0.4794	    \\
SPARISH 		& 0.6575          	& {\color[HTML]{3166FF} \textbf{0.6609}}		&{\color[HTML]{3166FF} \textbf{0.5110}}     \\
Synthetic-MaxPol	& 0.4873         	& 0.3656		&0.2824    \\
HVS-MaxPol  		& 0.5788         	& 0.5464		&0.4157    \\
FQPath 				& 0.5310         	& 0.5209		&0.3979    \\
FocusLiteNN 		& {\color[HTML]{3166FF} \textbf{0.6651}}          	& 0.5640		&0.4424    \\ 
HYQM     			& 0.3625         	& 0.2599		&0.2008    \\ \midrule
NSS-SR   			& 0.4293            & 0.3478		&0.2558          \\
LNQM     			& 0.6633          	& 0.6413		&0.4606		 \\ 
BSRIQA   			& 0.6587       		& 0.5639           &0.3987  	\\
DISQ 	    		&{\color[HTML]{CB0000} \textbf{0.7615}}	& {\color[HTML]{CB0000} \textbf{0.7643}}   	&{\color[HTML]{CB0000} \textbf{0.5784}}\\ \bottomrule    
\end{tabular}
\end{table}

%Overall, the DISQ outperforms the compared algorithms, which indicates that our method is a promising metric for SR-IQA. The superiority arises the aspect that our method fully utilizes related LR image as a reference, and fuses it into the regression network to provide more meaningful features for quality prediction, in order to improve the prediction performance.

\subsection{Ablation Study}

To evaluate the contribution of each component in the proposed DISQ method, we conduct a series of ablation experiments. The ablation results are presented in Table \ref{tab:reference} to \ref{tab:fusion}. In each column, the best results are highlighted in {\color[HTML]{CB0000} \textbf{red}}.

\subsubsection{Influence of the references LR images} In this experiment, we exclude the reference LR image in our model, resulting in a network with ${\rm CNN}_{\rm HR}$ and fully connected modules only. This no-reference model is trained on SISAR under the same parameter settings and training steps as the proposed reduced-reference DISQ model. Its testing performance on different databases is listed in Table \ref{tab:reference}. Although the performance of the no-reference model is slightly better on the QADS database, in all other three databases, the reduced-reference model performs much better, suggesting significant benefits of including the LR image as reference.
%In total, the proposed model with LR reference are evidently more promising than the simplified no-reference method, as can be evidenced from the higher PLCC value and better generalization performance. %The reason is that reference images provide extra informative features that are contributed to the network to learn the quality mapping function
\begin{table}[h]
\centering
\caption{PLCC Performance Comparison with or without LR image as Reference}
\label{tab:reference}
\setlength{\tabcolsep}{3.5mm}
\begin{tabular}{lcccc}
\toprule
Methods 	&  SISAR	& Waterloo-15  & CVIU-17 &QADS\\  \midrule
 w/o 		&0.8610	      	&0.6207	      	&0.5024 	&{\color[HTML]{CB0000} \textbf{0.8403}}	\\
 w/		&{\color[HTML]{CB0000} \textbf{0.9143}} 	&{\color[HTML]{CB0000} \textbf{0.8232}}      	&{\color[HTML]{CB0000} \textbf{0.6440}}		&0.8173	\\ \bottomrule    
\end{tabular}
\end{table}

It is also interesting to observe that the performance of this no-reference network is superior to many other algorithms shown in Table \ref{tab:comparison}. Experiments also reveal that further training of this no-reference model may increase its PLCC performance on SISAR but with a lower generalizibility to other databases. This suggests that including the LR image reference may help reduce overfitting.

\subsubsection{Selection of feature map pooling method} In the proposed network, the feature map is pooled into a concatenated tensor, which is described in Section \uppercase\expandafter{\romannumeral4}. \textit{B}. Previous studies \cite{Kang2014,Fang2018} have proved that the concatenated pooling feature has certain advantages compared with common mean pooling or max pooling. To intuitively illustrate the effectiveness of the joint features, we report the performance comparison of different pooling methods under identical training epochs in Table \ref{tab:pooling}, where the joint pooling method significantly improves the accuracy of quality prediction, especially on the CVIU-17 database. The merged feature map possesses rich and robust image features, which contributes to the mapping from image features to quality.
\begin{table}[h]
\centering
\caption{PLCC Comparison of different feature pooling methods}
\label{tab:pooling}
\setlength{\tabcolsep}{3mm}
\begin{tabular}{lcccc}
\toprule
Methods 	&  SISAR	& Waterloo-15  & CVIU-17 & QADS\\  \midrule
 Max Pooling 		&0.8803	      	&0.6593	      	&0.4512		&0.6714 	\\
 Mean Pooling		&0.8838	      	&0.7583	      	&0.4590 	&0.7712\\
 Joint Pooling		&{\color[HTML]{CB0000} \textbf{0.9143}} 	&{\color[HTML]{CB0000} \textbf{0.8232}}  	&{\color[HTML]{CB0000} \textbf{0.6440}}		&{\color[HTML]{CB0000} \textbf{0.8173}}\\ \bottomrule    
\end{tabular}
\end{table}

%%%%%%%%%%%%%%%%%%%

% We incorporate the LR and HR image features in different ways. In our reduced-reference network, the features extracted from the two CNNs are further fused to be the inputs of regression module. 

\subsubsection{Effectiveness of feature fusion methods} We incorporate the LR and HR image features in two other methods, which were also discussed in \cite{Bosse2018}. In summary, the image features $F_{\rm Hpool}$ and $F_{\rm Lpool}$ are merged in the following three methods %ways. Way 1 is utilizing the difference of these two features as the fused feature. Way 2 is the simplest way by concatenating $F_{\rm Hpool}$ and $F_{\rm Lpool}$. Way 3 directly connects the previous two approaches. 
\begin{equation*}
\begin{split}
&\textbf{Method 1: } F_{\rm fuse} = F_{\rm Hpool} - F_{\rm Lpool}, \\
&\textbf{Method 2: } F_{\rm fuse} = (F_{\rm Hpool}, F_{\rm Lpool}), \\
&\textbf{Method 3: } F_{\rm fuse} = (F_{\rm Hpool}, F_{\rm Lpool}, F_{\rm Hpool} - F_{\rm Lpool}). \\
\end{split}
\end{equation*}
%%%%%%%%%%%%%%%%%%%%%

Under the same training steps, the performance of our model combined with different feature fusion methods are provided in Table \ref{tab:fusion}. Method 1 exhibits the optimal performance with the fewest parameters on most databases. The network that merges $F_{\rm Hpool}$  and $F_{\rm Lpool}$ in Method 2 is theoretically capable of learning \begin{math} F_{\rm Hpool}-F_{\rm Lpool} \end{math} 
in the regression part. However, the performance is worse than Method 1, except for the QADS database. Method 3 unites the first two but fails to further improve the quality prediction accuracy even with an increasing number of paramaters. With better or similar PLCC performance, Method 1 uses only 1/2 and 1/3 of the numbers of parameters when compared with Methods 2 and 3, respectively.
%It shows that the explicit formulation like way 1 may simplify regression tasks and improve performance. Way 3 unite the first two ways cannot obtain higher accuracy, but increase the parameters. The reason may be that surplus features make the regression process more complicated and difficult. Even in the case of comparable performance, the parameters of way 1 are only $1/2$ of way 2 and $1/3$ of way 3.
\begin{table}[h]
	\centering
	\caption{PLCC of Models Using Different Feature Fusion Methods}
	\label{tab:fusion}
	\setlength{\tabcolsep}{2mm}
	\begin{tabular}{ccccc}
		\toprule
		%Fusion Manners 	& Shape of $F_{fuse}$   &  SISAR	& Waterloo-15  & CVIU-17 &QADS\\  \midrule
		\begin{tabular}[c]{@{}c@{}}Fusion Methods\\\& Shape of $F_{\rm fuse}$\end{tabular}  &  SISAR	& Waterloo-15  & CVIU-17 &QADS\\  \midrule
		Method 1: (3, 8, 8, 512)        &{\color[HTML]{CB0000} \textbf{0.9143}}      	&{\color[HTML]{CB0000} \textbf{0.8232}}    	&{\color[HTML]{CB0000} \textbf{0.6440}}	&0.8173\\
		Method 2: (6, 8, 8, 512)		& 0.8938       		& 0.7678			& 0.5626	&{\color[HTML]{CB0000} \textbf{0.8474}}\\
		Method 3: (9, 8, 8, 512) 		& 0.8973			& 0.7306			& 0.5214	&0.6897\\ \bottomrule    
	\end{tabular}
\end{table}

\section{CONCLUSIONS}
We build the largest of its kind SISAR database for SR-IQA, where a novel semi-automatic rating approach is proposed, which greatly reduces the workload for human labeling. We train an end-to-end DISQ model for SR-IQA which uses a two-stream DNN for feature extraction, followed by a feature fusion network for quality prediction. We train the DISQ model using the SISAR database, and the experimental results demonstrate the superior performance of the DISQ model against state-of-the-art. Cross-database validation also shows better generalizibility of the proposed DISQ model. The SISAR database and the DISQ model will be made public available to facilitate reproducible research.

\ifCLASSOPTIONcaptionsoff
  \newpage
\fi



%\bibliographystyle{IEEEtran}
\balance
\bibliographystyle{IEEEtran}
\bibliography{./ref1}%

\end{document}